\documentclass[useAMS,usenatbib]{mn2e}
\usepackage{epsfig,appendix,times,graphics}
\usepackage{color}

\newcommand{\al}{et al. }
\newcommand{\hi}{H\,{\sc i}}
\newcommand{\hii}{H\,{\sc ii}}
\newcommand{\prim}{$^{\prime}$}
\newcommand{\prin}{$^{\prime\prime}$}
\newcommand{\aprox}{${\sim}$}

\newcommand{\km}{~km~s$^{-1}$}
\newcommand{\degree}{$^{\circ}$}
\newcommand{\halpha}{H${\alpha}$}
\newcommand{\cg}{CGCG\,097-}
\newcommand{\msolar}{M$_{\odot}$}

\newcommand{\defhi}{H{\sc i} deficiency}

\newcommand{\first}{The velocity in \km\ is shown in the top righthand corner of each frame. Contour levels are at -2, 2, 3, 4, 6,10, 15, 20, 30 and 45 $\sigma$ where  $\sigma$  corresponds to }
\newcommand{\last}{The cross marks the centre of the optical galaxy. }

\title[]{Probing Evolutionary Mechanisms in Galaxy Clusters: Neutral Atomic Hydrogen in Abell\,1367}
\author[]{T. C. Scott$^{1}$\thanks{E-mail:
t.c.scott@herts.ac.uk (TS)}, H. Bravo--Alfaro$^{2}$, E. Brinks$^{1}$, C. A. Caretta$^{2}$,
L. Cortese$^{3}$, A. Boselli$^{4}$ \and M. J. Hardcastle$^{1}$, J. H. Croston$^{1}$ and I. Plauchu$^{2}$
\\
$^{1}$Centre for Astrophysics Research, University of Hertfordshire, College
Lane, Hatfield, AL10 9AB, UK\\ 
$^{2}$Departamento de Astronom\'\i a, Universidad de Guanajuato.
Apdo, Postal 144, Guanajuato 36000, Mexico\\
$^{3}$School of Physics and Astronomy, Cardiff University, Cardiff CF24 3AA,
UK \\ 
$^{4}$Laboratoire d'Astrophysique de Marseille, OAMP, Universit$\acute{e}$ Aix--Marseille 8 CNRS, 38 Rue Fr$\acute{e}$d$\acute{e}$ric Joliot--Curie, 13388 Marseille, France}
\begin{document}
\date{Accepted . Received ; in original form }


\maketitle

\label{firstpage}

\begin{abstract}
\noindent
We present VLA \hi\ imaging data for a field in the NW of the galaxy cluster Abell 1367 (z\,=\,0.02) in an attempt to probe the effect environment has on the interstellar medium of late--type spiral galaxies. Several galaxies, like CGCG 097--087, show pronounced tails and asymmetries, and 7 out of 10 show significant, several kpc offsets between the HI centroid and the optical. We compare our results against a sample of optically bright, late--type galaxies (spirals) across the central 1.5 Mpc of the cluster taken from the {\it Arecibo Galaxy Environment Survey} (AGES). We find that these late--type spirals are predominantly found in the northern half of the cluster, especially those that are relatively gas--rich. We calculate the \hi\ deficiency and find that the expected global trend for the \hi\ deficiency of these spirals to increase with projected proximity to the cluster core, seen in clusters like Coma and Virgo, is not observed. We classified the spirals into four evolutionary states, with the galaxies in each state sharing a similar degree of \hi\ deficiency and optical colour. The common characteristics of the spirals in each evolutionary state suggests they have been subject to similar environmental processes. Many of the spirals in the most common evolutionary state (moderate \hi\ deficiency and blue colour) have an \hi\ intensity maximum which is displaced relative to its optical counterpart. The orientation of these offsets and magnitude of their \hi\ deficiencies together with data from other wavelengths provide observational evidence in support of varying degrees of ram pressure stripping and tidal interaction. In general, our results indicate that the \hi\ disks of bright late--type galaxies in the central part of the cluster are subject to both gas loss and morphological disturbance as a result of their interaction with the cluster environment. This provides further observational evidence of a more complex environment in Abell 1367 as compared to Virgo and Coma.

\end{abstract}


\section{INTRODUCTION}

For several decades it has been known that the fraction of spiral galaxies
decreases when moving from a field population to a galaxy cluster core, with
a corresponding increase in the fraction of early type (E+S0) galaxies; this
is known as the morphology--density relation
\citep{dress80,oemler74}. Whether this relation originates during galaxy
formation or is mainly the result of environment remains one of the
fundamental questions in observational cosmology. While previous work
\citep[Boselli $\&$ Gavazzi 2006a;][ and references therein]{dress99,poggi99,dress04} \nocite{bosel06a}suggests that the
environment associated with clusters only plays a secondary role in producing
lenticular galaxies from spirals, there is ample evidence for the effect of
cluster environment on galaxy evolution, such as the observed increase in the
fraction of \hi--deficient spiral galaxies toward cluster cores 
\citep[e.g., Boselli $\&$ Gavazzi 2006a;][ and references therein]{vgork04}, \nocite{bosel06a} where the \defhi\ is defined as the log of the difference between the expected and 
observed \hi\ mass (see Table~\ref{sdss}, table footnote $h$). There is strong evidence that infalling cluster galaxies encountering a hot X--ray emitting intracluster medium (ICM) for the
first time experience rapid evolution as a result
\citep[Gavazzi \al 2001b;][]{sun02,ken04,vgork04}.\nocite{gava2001b} In these cases the interaction removes the interstellar medium (ISM) from the galaxies \citep[e.g.,][]{gava95,bravo00,sola01}. For infalling spirals the gas motions resulting from the interaction are expected to lead to quenching of star formation. Although modest and short timescale (10$^8$ years) increases in star formation are predicted in specific circumstances \citep{fujita99,kronb08}. However the predicted ram pressure induced star formation was not observed in a study by \cite{iglesias04}.

Different types of mechanisms are proposed to explain how the cluster
environment impacts evolution of its constituent galaxies. The first
type, which is most evident in spirals, consists of hydrodynamic mechanisms
arising from gas--phase interactions between the cluster's hot ICM and a
galaxy's ISM; these include ram pressure stripping
\citep{gun72}, viscous stripping \citep{nuls82}, thermal evaporation \citep{cowie77} and starvation
\citep{bekki02,fuji04}. The second type is caused by gravitational
effects produced either by close encounters with neighbouring galaxies
(merging or tidal stripping) or by repeated high speed galaxy--galaxy
encounters  \citep[{\it harassment;} ][]{moore96,moore99}. Also in this
category belongs gravitational interaction between a galaxy and the
cluster potential as a whole \citep{bekki99,nata02}.

\hi\ in spirals provides an excellent tool to investigate these mechanisms as
it is sensitive to both gravitational and hydrodynamic interactions
\citep{dick91}. There is conflicting evidence for and against a correlation between the fraction of \hi\ deficient galaxies and  the cluster X--ray luminosity, $L_X,$  \citep[Boselli $\&$ Gavazzi 2006a;][]{giovan85,sola01}. \nocite{bosel06a} However modelling by \cite{tonn07} favours ICM stripping over gravitational effects as the primary cause of \defhi\ in nearby cluster spirals. Modelling of a generic spiral infalling to a cluster predicts the removal of almost all of its \hi\ during the first transit of the cluster core as a result of hydrodynamic stripping. During the transit a combination of high ICM density and maximum orbital velocity hugely increases ram pressure stripping efficiency
\citep[Vollmer  et al 2001a;][]{roed07}\nocite{voll01a}. The few individually modelled spirals in Virgo,  e.g., NGC 4522 \citep{ken04,voll08}, NGC 4569 \citep[Boselli et al.2006b;][]{voll04} \nocite{bosel06b} and in Coma, NGC4848 (Vollmer \al 2001b)\nocite{voll01b}, confirm that the ram pressure mechanism is operating. There are cases in Virgo, e.g., NGC4654 \citep{voll03}, NGC 4438 \citep{bosel05}, NGC 4254 \citep{chy08} and \citep{chung07} as well as examples in other clusters \citep{sun05b,moran07}  indicating that tidal mechanisms are operating in addition to hydrodynamic mechanisms.

The issue of galaxy evolution near clusters has become richer and more
complex in recent years, since modelling suggests that under the hierarchical
large--scale structure formation scenario, clusters are expected to grow
predominantly by accretion of groups of galaxies rather than individual
galaxies(Blummenthal et al. 1984; Springel et al. 2001, although for an alternative model see Berrier et al. 2009) \nocite{berri09,blum84, spring01}. Infalling groups, having lower velocity dispersions than the
clusters, provide opportunities for tidal interactions which may represent an
important {\it preprocessing} stage in cluster galaxy evolution
\citep{mihos04,fuji04,dress04}.  \cite{wilman09} report strong evidence that transformation of spirals to S0s in clusters occurs  preferentially in such groups  as a result of tidal interactions at z $\sim$0.4. Their results are based on analysis of HST-ACS images of 179 spectroscopically confirmed group and 111 field galaxies from the Canadian Network for Cosmology (CNOC2) sample. This raises the question of what processes have perpetuated the transformation of late--type galaxies to S0s since then \citep{fasa00}, given in the current epoch hydrodynamic mechanisms appear to be the predominant interaction type, but the most likely mechanisms for conversion of spirals to S0s are widely thought to be tidal \citep{dress80,bick02,wilman09}. 

In this paper we consider A\,1367 which together with A\,1656 (the Coma cluster) comprises the Coma supercluster. A\,1367 ($z$=0.022) lies at the intersection of two filaments; the
first extending 100 Mpc eastwards in the direction of the Virgo cluster and
the second running NE toward Coma \citep{west00}. A\,1367, with a
mean velocity $\approx$ 6,240\,\km\ and \ ${\sigma}$ $\approx$ 822\,\km, is
an unrelaxed Bautz--Morgan type II--III, spiral rich, 6.9 x
$10^{14}$\msolar\, cluster (Boselli $\&$ Gavazzi 2006a) \nocite{bosel06a}. Based on a redshift to the
cluster of 0.022 and $\Omega_M =0.3$ , $\Omega_\Lambda =0.7$ \ and \ $H_0$ =
72 \km\ Mpc$^{-1}$ the angular scale is  1 arcmin = 24.8 kpc.

A\,1367 is, for several reasons, the perfect target to study environmental
effects on spirals. It has a much higher fraction of spirals than its more
massive neighbour Coma. X--ray observations of A\,1367 carried out with
ASCA, \textit{XMM--Newton (XMM)} and {\em Chandra} indicate a dynamically young system with
two principal sub--clusters \citep[e.g.,][]{don98}. These two sub--clusters are themselves in the process of assembling from
several smaller groups \citep{cort04}. This suggests that A\,1367 may be more like the unrelaxed clusters at earlier epochs than the more relaxed systems typical of low redshift  \citep{oemler97}. Being a lower mass cluster of the kind in which \cite{pog09} report enhanced rates of morphological transformation from spirals to S0s it is potentially a location in which to test the mechanism leading to galaxy transformation suggested by  \cite{wilman09}. Furthermore,
extensive optical and radio continuum studies have been carried out for
A\,1367. Gavazzi, in a series of papers \citep[Gavazzi \al 2001a;][]{gava99,gava95,gava87} \nocite{gava2001a}
reported several galaxies with radio continuum and \halpha\ tails and
exceptionally high star formation rates NW of the cluster centre, interpreted
as arising from the interaction of galaxies with the cluster ICM. Also the
recently discovered Blue Infalling Group (BIG), projected near the centre
of A\,1367 with a high relative velocity (Sakai \al 2002; Gavazzi \al 2003a; Cortese \al 2006) \nocite{sakai02,gava03a,cort06},
may constitute a unique laboratory for testing {\em preprocessing} theories in cluster assembly.

This paper is the first stage of an investigation into the mechanisms
transforming late--type galaxies in Abell 1367 which is focused on their \hi\
content and position within the cluster. It is based on both Arecibo\footnote{ The Arecibo Observatory is part of the National Astronomy and Ionosphere Centre, which is operated by Cornell University under a cooperative agreement
with the National Science Foundation
} single dish and NRAO\footnote{The National Radio Astronomy Observatory is a facility of the National Science Foundation
operated under cooperative agreement by Associated Universities, Inc.}-VLA
synthesis \hi\ imaging. Despite Arecibo's modest spatial resolution of \aprox
3.5 arcmin, its homogeneous (in \hi\ sensitivity) sky coverage, good sensitivity and high velocity
resolution are well suited to study the global \hi\ distribution throughout
nearby clusters. But interferometers, like the VLA, provide an order of magnitude better spatial resolution, and for nearby clusters the VLA's spatial resolution is sufficient to  allow the effects of environment on the \hi\ distribution in individual galaxies to be studied.

In this paper we present the results of a VLA low/medium resolution \hi\
spectral line study of two adjacent fields in the central and NW parts of
A\,1367. The \hi\ imaging from these two fields covers a small fraction of the virial volume of the cluster.

In section 2 we give details of the VLA \hi\ observations including data
reduction. Special attention is given to the method we developed to optimise
the continuum subtraction, leading to improved signal-to-noise for low
brightness \hi\ features. We report our main observational results from the VLA \hi\ imaging of individual galaxies in section 3
and, in section 4 the distribution of \hi\ throughout the cluster volume from the AGES data. In
section 5 we discuss  the implications of our data for evolutionary
mechanisms in the cluster, based on the \hi\ content and colour distributions at
both large and small scales as well as on the peculiar \hi\ morphologies
shown by several galaxies. Our concluding remarks are given in
section {\ref{conclusion}}.

\section{OBSERVATIONS}
\label{obs}

\begin{figure}
\begin{center}
\includegraphics[scale=0.4] {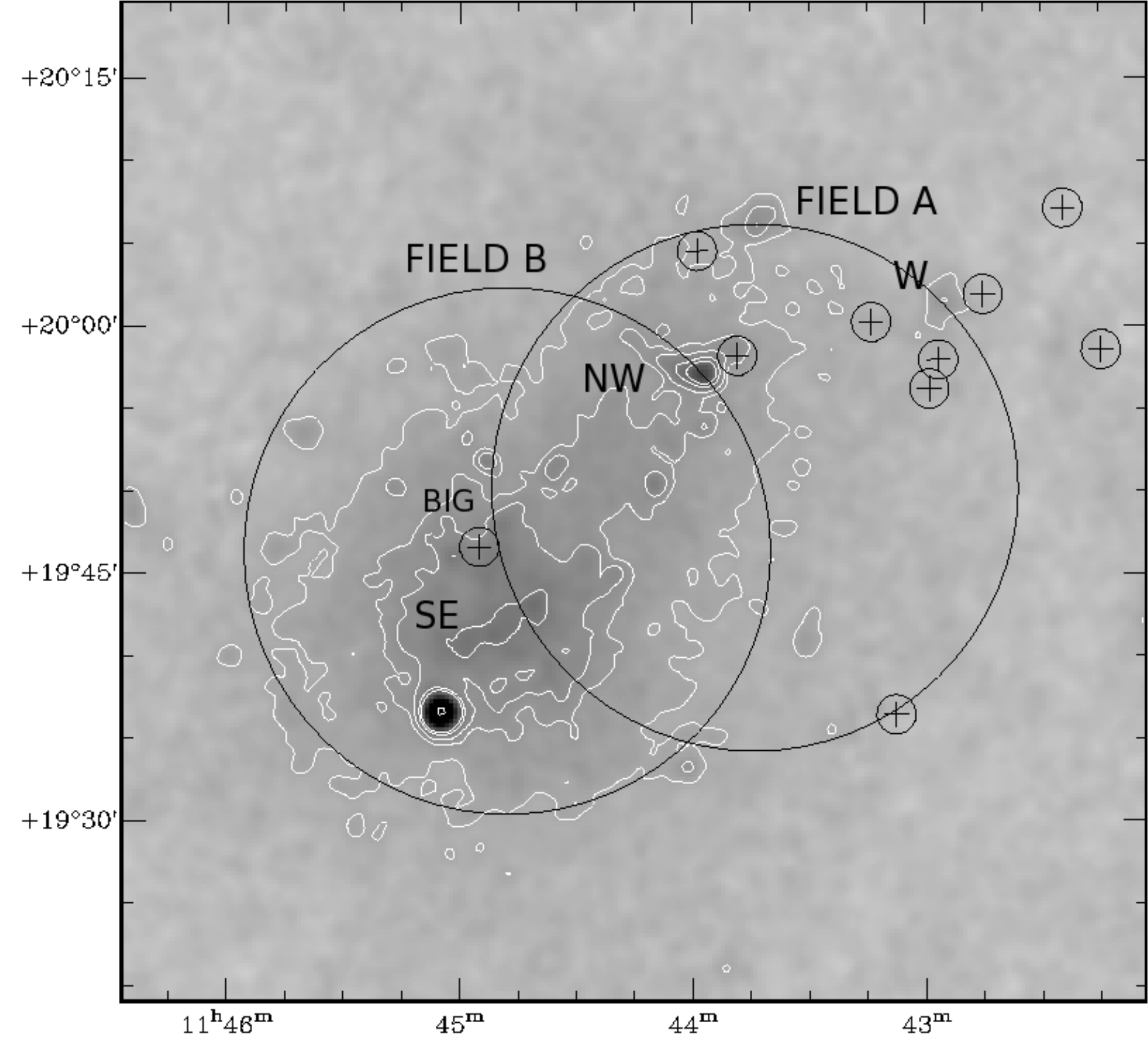}
\caption{The positions of the two fields observed with the VLA are
indicated by large circles showing the FWHM (32 arcmin) primary
beams. VLA \hi\ detections are shown with small black circles with crosses. The grayscale is a \textit{ROSAT} X--ray image with white contours overlaid. The positions of the SE, NW and W
sub--clusters and the BIG compact group are labelled }
\label{bigrosat} 
\end{center}
\end{figure}

We used two sets of VLA \hi\ observations both of which are well within the
104 arcmin (2.58 Mpc) virial radius of the cluster \citep{moss06} and
completely within the A\,1367 AGES surveyed volume \citep{cort08}. Figure \ref{bigrosat}
indicates the FWHM primary beams for both VLA pointings (large circles). The
figure also shows the intensity of the X--ray emission {\em (ROSAT)} from the
cluster's ICM (white contours) and the positions of the VLA \hi\ detections
(small black circles enclosing crosses). Observational parameters for both fields are listed in
Table \ref{tt1}, including the velocity resolution, central velocities, and
the rms noise per field.

The sensitivity is non-uniform between the two VLA fields because of their differing array configurations and integration times. Additionally, for both fields, there is a sharp drop in
sensitivity beyond the primary beam FWHM radius, significantly impacting the three
detections in Field A furthest from the pointing centre.

\begin{table*}
\centering
\begin{minipage}{140mm}
\caption{VLA Observational Parameters}
\label{tt1}
\begin{tabular}{@{}ccccccccccc@{}}
\hline

Field \footnote{Field A = VLA Project ID: AB900; Field B = VLA project ID AH801}&
${\alpha}_{2000}$ & ${\delta}_{2000}$ & Array &Integration & Beam &
Velocity & Central & rms noise & mJy  \\

& & & config & time &size & resolution & velocity & 
& conversion \\
& [$^h$ $^m$ $^s$] & [\degree\ \prim\ \prin\ ] & & [hours]& [\prin]&
[\km ] & [\km ] & [mJy\,beam$^{-1}$ ]&[K]\\ \hline A & 11 43 45.6 & 19 50 21
& D &15.0& 44 x 45 & 21 & 6800 & 0.27 &0.3 \\ B &11 44 50.0 & 19 47
00 & C & 3.6 & 15 x 15 &11 & 8200 & 0.65& 2.7 \\
 \hline
\end{tabular}
\end{minipage}
\end{table*}

\subsection{\bf VLA Field A - NW sub--cluster}
\label{vlafielda}
Field A is centred near the NW sub-cluster \citep{cort04},
\aprox10 arcmin NW of the cluster centre (position from NED between the NE and SE sub--clusters). Field A's position was chosen based on two considerations. First, the peculiar Zwicky infalling starburst
galaxies \cg073, \cg079, and \cg087 were imaged within the 32 arcmin FWHM of Field A's primary beam. Second, the first null of
the beam was placed near 3C264 (a strong continuum source, with a flux density of 5 Jy at 1.4 GHz) to minimise the effects of its sidelobes.

This field was observed using the VLA in D configuration on the 26th and 30th
of March 1999. Total observing time for the field was 15 hours. We used correlator mode 4 which gives two independent Intermediate Frequencies (IFs), each with dual polarization (right and left-hand circular). The IFs for the first day were set so that each IF
generated a position--position--velocity subcube ($\alpha, \delta, $velocity) with a velocity range of \aprox 500 \km, with a small
overlap between the velocity ranges. Combining the subcubes from both IFs
produced a cube for the field with a velocity range of \aprox 1000 \km. The
field was re--observed on the second day using the same procedure but for a range
of velocities adjacent to that observed on the first day. When combined, the
subcubes from both days produced a single cube of 96 channels
with a continuous velocity range from 5762 to 7810\,\km\ and a velocity
width, using on--line Hanning smoothing, of 21\km\ per channel. The final
cube's velocity range is more than twice the velocity dispersion of A\,1367
(\aprox822\,\km).

The data were calibrated and imaged following standard procedures using the AIPS
software package. For Field A, self--calibration with a single iteration was carried
out to improve upon the standard complex gain calibration. This was necessary
to mitigate the effects of sidelobes from 3C264. We applied different robust
weighting functions in the Fourier transform (AIPS task {\sc imagr}),
looking for a compromise between sensitivity and resolution. Our final data cubes were produced with robust weighting (using {\sc robust = 0}) and have a resolution of about 45 arcsec. The {\sc robust } option corrects the weights of the visibilities in the Fourier transform for the fact that there is a much higher density of measured visibilities in the inner part of the {\em uv--}plane compared to the outer regions \citep{briggs95}. This comes at a cost of a slightly increased noise compared to using what is known as 'natural weights'. For this data set the \hi~mass detection threshold is
\aprox 7$\times$10$^{7}$ M$_{\odot}$ (corresponding to 3$\sigma$ over 2
consecutive 21 km~sec$^{-1}$ channels). The equivalent column density
sensitivity for emission filling the beam is then 1.9$\times$10$^{19}$
cm$^{-2}$.

It was found that continuum subtraction was a critical step in the data
reduction. Ordinarily one would search for line--free channels in each
subcube and then use, e.g. {\sc uvlin} to subtract the continuum. The problem
with our observations was that there was some \hi\ emission at some position in almost every channel. This required a more elaborate approach consisting of deriving a continuum map made up of those areas on each channel map with no line emission and averaging them. This map was subsequently subtracted from the line + continuum data to produce cubes containing only line emission. In practice this required the following procedure. First, continuum sources with peak flux density $\geq$ 10 mJy were modelled and removed in the {\em
uv}--plane. This resulted in subcubes containing line and residual continuum
sources of $\leq$ 10 mJy. In the second stage AIPS tasks {\sc sqash} and {\sc comb}
were used to create an average of all channels from the line plus residual
continuum cube in the image plane, which was subtracted from the line plus
residual continuum cube. This enabled us to find line--free channels in the subcubes, although in most cases there were very few (just
1 or 2). Continuum subtraction was then improved by
repeating the subtraction but only using the average of these few line free
channels to produce subcubes with \hi\ line emission only. The rms noise in
these cubes was quite high because the continuum subtracted was
based on very few channels.

In the final stage these \hi\ cubes were smoothed and blanked using AIPS tasks {\sc convl} and
{\sc blank}, blanking out all areas in each channel containing line emission. These
masks were in turn applied to the \textit{line + residual} continuum subcubes
to create, after again applying {\sc sqash}, residual continuum--only maps for each of the subcubes. These residual continuum maps were then subtracted from the
original {\it line + residual} continuum subcubes, resulting in subcubes with
only \hi\ emission, but with a much reduced rms noise. The continuum
subtracted subcubes were subsequently combined with {\sc mcube}. The final cube produced in
this way had a noise of \aprox 0.27 mJy\,beam$^{-1}$ which increased the
number of \hi\ detections by 50 percent compared to the equivalent cube produced
using {\sc uvlin}.

\subsection{\bf VLA Field B - BIG} 
The second set of VLA \hi\ observations was centred on the Blue Infalling
Group (Figure \ref{bigrosat}), overlapping slightly and located to the SE of
Field A, with a central velocity of 8200\,\km, i.e. the mean velocity of the BIG
galaxies. Archival C--array observations (correlator mode 2 with on--line Hanning
smoothing) from 8 December 2002 were used to produce an image cube with 48 channels
(after discarding the noisy edge channels) with a velocity width of 11 \km\
per channel. The velocity coverage was 7913--8587 \km. Our final data cube has a
resolution of \aprox 15 arcsec. In the spatial overlap region between the two
observed VLA fields, the velocity range extends from 5762 to 8587 \km\
(except for a narrow gap at \aprox 7900 \km). For Field B, the equivalent
\hi~mass detection threshold is \aprox 8 $\times$ 10$^{7}$ M$_{\odot}$
(corresponding to 3$\sigma$ in 2 consecutive 11 \km\  channels).  This
is equivalent to column density sensitivity for emission filling the beam
of 2.1 $\times$ 10$^{20}$ cm$^{-2}$.

\section{OBSERVATIONAL RESULTS}

\subsection{The \hi\ in individual galaxies: VLA observations}
Nine objects were detected in Field A \citep[NW and W sub--clusters,][]{cort04} and two more
within Field B (BIG). Channel maps for each detected object are available in Appendix C  online material. Figure \ref{bigxmm} shows the VLA \hi\ detections in
the NW area of the cluster, and their locations relative  to
X--ray emission (from {\em XMM--Newton}). The {\it XMM-Newton} archive data were derived from an observation taken on 2001 Nov 22. After standard flare filtering, the exposure
time was 20.6 ks for the pn and 29.9 ks for the two MOS cameras.
Images in the 0.5-5.0 keV energy band were derived by combining the
exposure-corrected images from all three cameras and convolving with a
Gaussian. 

The morphological type, EW(\halpha), VLA \hi\ velocity  including uncertainty,    ${\Delta}$V$_\mathrm{HI}$ including uncertainty, \hi\ mass, and offset
(of the \hi\ compared to the optical position)  for the VLA detections are given in
Table \ref{table2}. 

The uncertainty in position for sources in the VLA fields is \aprox\ 4 arcsec (1.6\,kpc) and 2 arcsec (0.8\,kpc) for Field A and B respectively, i.e, uncertainties are $\sim$ 1/10th of the synthesized beam. Within a velocity integrated intensity map of an individual spiral, positional uncertainly varies depending on S/N. This means positional uncertainty within a map of a spiral is lowest in the region of velocity integrated intensity maximum. But  we caution that the distribution beyond the \hi\ intensity maximum region in spirals beyond the FWHM of the primary beam is more uncertain.

Four galaxies, \cg062, \cg068, \cg072 and \cg125, imaged with the VLA  have their \hi\ intensity maximum offset relative to their optical counterpart (final column of Table \ref{table2}) in directions which are not radial with respect to the cluster centre, i.e. these cases are inconsistent with a simple ram pressure scenario. The direction of these offsets may arise from a tidal interaction or a more complex ram pressure scenario such as gas fall--back (see subsection 3.3 for \cg072 below).

\begin{table*}
\centering
\begin{minipage}{140mm}
\caption{Parameters for the VLA \hi\ detected galaxies}
\label{table2}
\begin{tabular}{|l|l|l|l|r@{}   |r@{}  |l|l@{} |r@{}|l@{}|l@{}|l@{} |l} \hline 
ID\footnote{Zwicky catalogue, except GP1277 = Abell 1367[GP82]1227 and GP1292 = Abell 1367[GP82]1292}
&$\alpha_{2000}$\footnote{Optical galaxy position from NED} 
&$\delta_{2000}$ 
&Type\footnote{Hubble classification from NED, * = our classification}
& EW(\halpha)\footnote{ EW(\halpha) is from GOLDMine, except GP1227 which is our measurement}& V$_\mathrm{HI}$\footnote{Velocity = the mean of the upper and lower velocity defined by the line width  (see note \textit{g}) }
& $\sigma$(V$_\mathrm{HI}$) \footnote{\ Uncertaintiy for V$_\mathrm{HI}$ = 1.5 (W$_{20}$--W$_{50}$)(S/N)$^{-1}$ \citep{schnei90}, except where W$_{20}$ = W$_{50}$ in which case $\sigma$(V$_\mathrm{HI}$) = 1.5 (2 $\times$ channel width )(S/N)$^{-1}$.  }
&W$_{20}$& $\Delta$V$_\mathrm{2 \sigma }$\footnote{The velocity width based on the number of contiguous channels with  $\geq$2 $\sigma$  detections.}
& $\sigma$($\Delta$V$_\mathrm{2 \sigma }$)\footnote{Uncertainty in the line width = 2$\sigma$(V$_\mathrm{HI}$) }
&M$_\mathrm{HI}$\footnote{M$_\mathrm{HI}$ = 2.36 $\times$10$^2$ D$^2$S$_\mathrm{HI}$ where M$_\mathrm{HI}$ is in \msolar, D in Mpc and S$_\mathrm{HI}$ is  the VLA flux in mJy \km. Mass based on a distance of 91.6 Mpc ($V_{A\,1367}$ = 6595\km\ from NED and $H_{0}$ = 72\km\ Mpc$^{-1}$.) }

& \multicolumn{2}{c}{Offset \footnote{The offset direction is the projected position of the \hi\ intensity maximum relative to its optical counterpart, except for \cg087 where the direction and distance are taken from the NIR counterpart. The projected distance between the two intensity maxima is given in arcsec with a positional uncertainty of \aprox\ 4 arcsec. }} \\ 
 &&&&&VLA&&&VLA&&VLA\\
& [ $^h$ $^m$ $^s$ ] &[\degree \prim\ \prin] && [\AA] & [\km]& [\km]&[\km]&[\km]& [\km] &[$10^{9}$
\msolar] &\\
\hline
97-068 & 11 42 24.5 & 20 07 10 &Sbc & 41 &  5958 & $\pm$\,\,5& 346& 346&$\pm$10  & 6.9 &15\prin & N \\
GP1227 & 11 43 13.0 & 19 36 47 &Dw* & 30 & 6239 & $\pm$11&  86& 86&$\pm$22& 0.9 &none &  \\
97-072 & 11 42 45.2 & 20 01 57 &Sa & 9 & 6207 &$\pm$\,\,4 &  64&64&$\pm$\,\,9& 0.4 &12\prin & SE \\
97-087 & 11 43 49.1 & 19 58 06 &Im & 77 &  6738 &$\pm$27&  563 & 649 &$\pm$54& 8.0 &30\prin & NW \\
97-079 & 11 43 13.4 & 20 00 17 &Irr &129&  7019 &$\pm$21&  216&216&$\pm$41& 1.3 &10\prin & NW \\
GP1292 & 11 42 58.9 & 19 56 12 &Dw* & - &  7203 & $\pm$23&  64& 64&$\pm$46& 0.3 &none& \\ 
97-073 & 11 42 56.4 & 19 57 58 &SA:pec &111& 7301 &$\pm$11&  173&216 &$\pm$22& 2.0 &8\prin & N \\
97-091 & 11 43 59.0 & 20 04 37 &Sa & 23 &  7377 &$\pm$\,\,2&  259& 281&$\pm$\,\,4& 5.3 &none &\\
97-062 & 11 42 14.8 & 19 58 35 &Sa:pec & 37 &  7723& $\pm$13&  64& 64&$\pm$26 & 0.7 & 15\prin & SW \\
97-125 & 11 44 54.8 & 19 46 35 &S0a & 23 &  8158 & $\pm$18&  206& 206&$\pm$35& 1.3& 12\prin &SW \\
K2 & 11 44 50.6 & 19 46 02 &\hii\ & - & 8158&$\pm$12&  162&162&$\pm$24& -  &-& - \\
\hline
\end{tabular}
\end{minipage}
\end{table*}

Most of the galaxies have previously been detected in \hi\ with Arecibo
\citep{chincar83,giovan85,gava89,springo05} and/or with the VLA
\citep{dick91,hota07}. None of the authors showed \hi\ maps of their
detected galaxies because of insufficient resolution or sensitivity, except
\cite{hota07} who presented \hi\ maps of the three most active starburst
galaxies. 

All objects in Table \ref{table2} were also detected in AGES
\citep[Table \ref{sdss} and][]{cort08}, with total fluxes and velocities
generally in good agreement with ours. As an example we show, in Figure
\ref{87spec}, a comparison of the AGES and VLA Field A spectra for a detection
well within the VLA FWHM beam. Important exceptions to this agreement
include \cg073 and A\,1367 [GP82]1292, which are confused in the AGES beam
but are resolved by the VLA. 

While the determination of velocity for each channels is accurate to within a few \km\ the range of velocities detected for each galaxy depends on sensitivity. The VLA detections of \cg068, \cg072 and \cg062
are beyond the Field A primary beam, where not only is the sensitivity lower
than in AGES, but the fluxes are more uncertain due to the limited knowledge
of the VLA primary beam beyond the half--power radius.  For example the noise at the position of \cg068 is $\sim$5 times greater than at the centre of the primary beam. 

In the following subsections we describe the most notable
features of the VLA detections. The images displayed here were produced
after applying a primary beam correction. The $\alpha$ and $\delta$ axes of all maps are in J2000 coordinates.

\begin{figure}
\begin{center}
\includegraphics[scale=0.48] {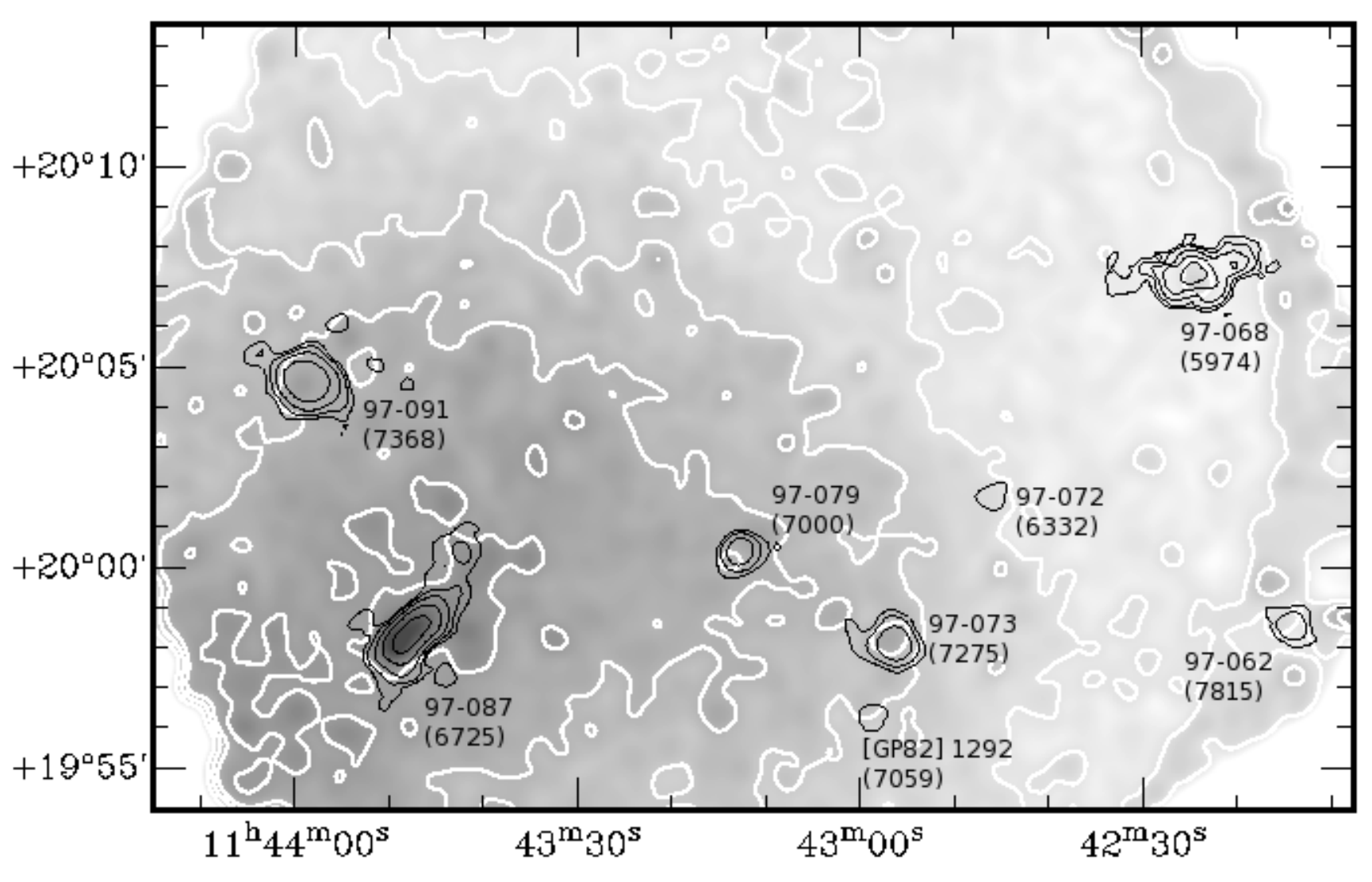}
\caption{\textbf{NW region of A\,1367: }X--ray emission (XMM) image with white
contours. Black contours indicate \hi\ emission from the VLA \hi\ detections in field A, with the outer contour indicating a column density of \textit{N}$_{HI}$ = 3
x 10$^{19}$ cm$^{-2}$. Values in parentheses below the Zwicky identifier are optical velocities in \km.}
\label{bigxmm} 
\end{center}
\end{figure}

\begin{figure}
\begin{center}
\includegraphics[scale=0.6] {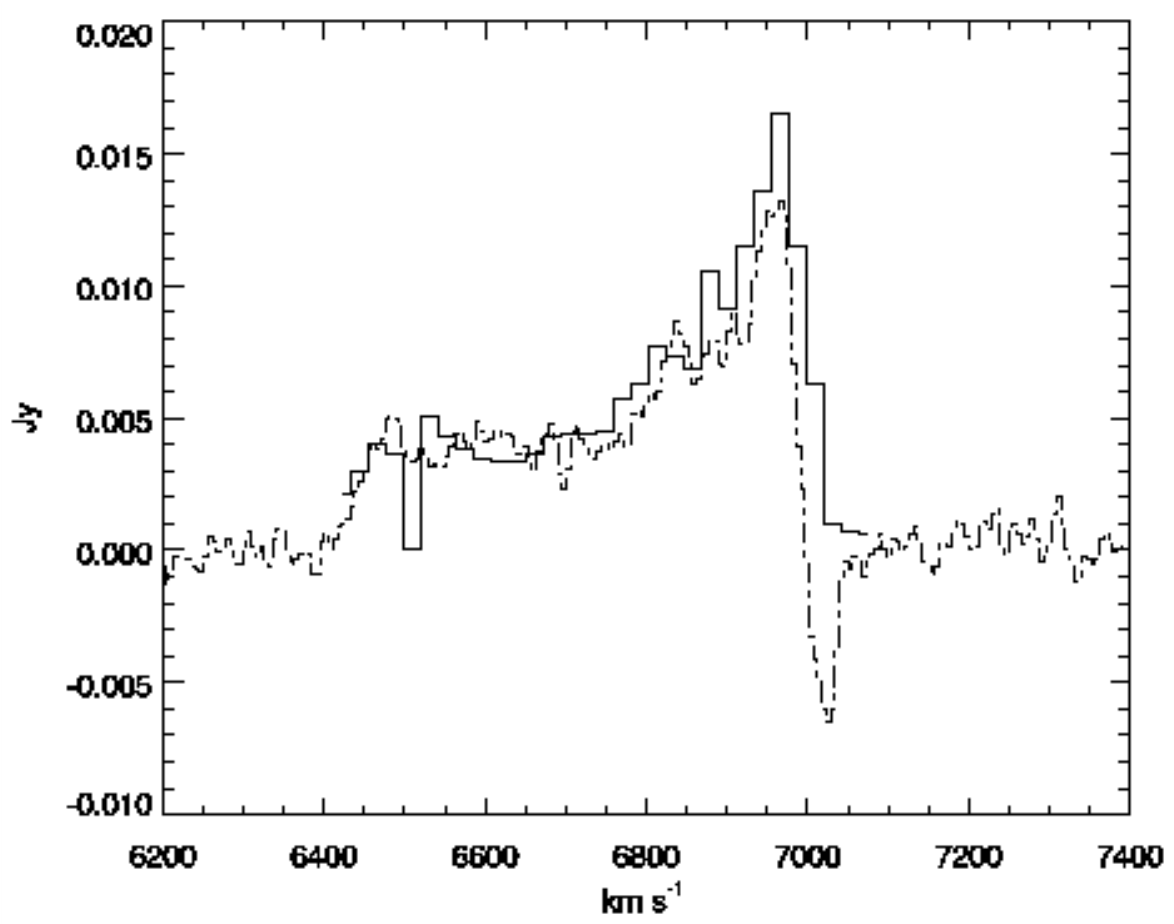}
\caption{\cg087 VLA (solid line) and AGES (dashed line), integrated
\hi\ spectra. The feature at 7000 \km\ in the AGES spectrum is an RFI
artifact.}
\label{87spec} 
\end{center}
\end{figure}

\subsection{ \bf The star-forming galaxies (\cg087, \cg073 and \cg079)}
\label{starbusts}

These three starburst galaxies have been extensively studied in the
past \citep[Gavazzi \al 2001a,b;][]{nuls82,gava84,gava87,gava89,gava95,dick91,bosel94,hota07}\nocite{gava2001a,gava2001b}. They all have exceptionally high star formation rates, EW(\halpha\ + [NII])$>$ 75\AA\ (Table \ref{table2}) and display spectacular radio continuum and \halpha\
tails pointing away from the cluster centre  \citep[Gavazzi \al 2001b;][]{gava87} \nocite{gava2001b}. Some
observational features of these three objects have been interpreted as
effects of ram pressure stripping produced by the ICM, in particular the
enhanced star formation activity and the displacement of atomic hydrogen
relative to their stellar disks, although in the case of \cg087
much of the enhanced star formation rate is probably attributable to a merger
 \citep[Gavazzi \al 2001a;][]{mart08} \nocite{gava2001a} The \hi\ in the central parts of these starbursts
has been recently mapped by \cite{hota07}. The maps presented here, though,
show additional large scale \hi\ features as a result of our improved
continuum subtraction (see Sect. \ref{vlafielda}).

\begin{figure}
\begin{minipage}{\linewidth}
\centering
\includegraphics[scale=0.5] {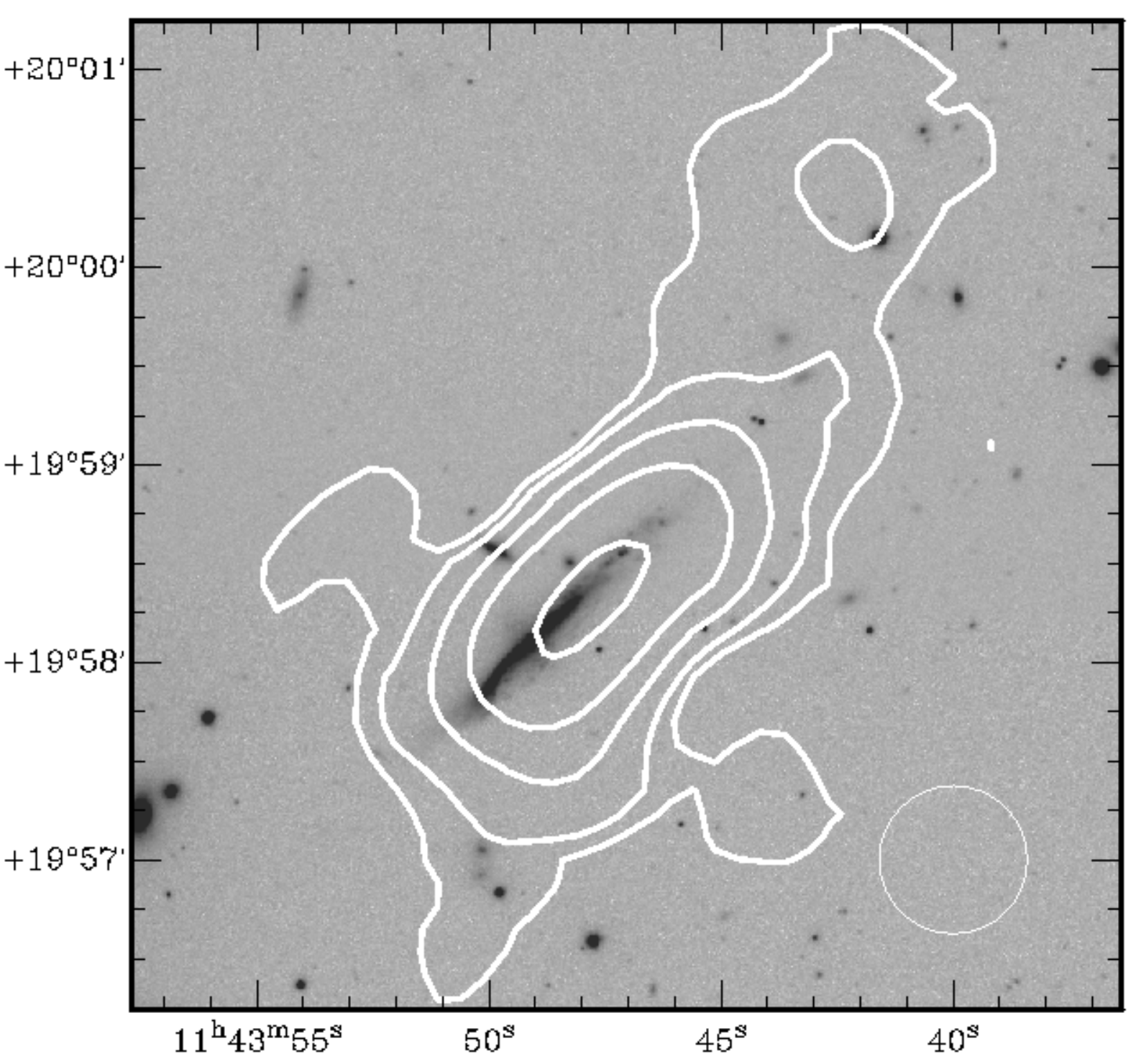}
\caption{\textbf{\cg087:} White contours are from a robust 0 \hi\
surface density map, with the outer contour indicating a column density of  \textit{N}$_{HI}$  = 3
x 10$^{19}$ cm$^{-2}$, with higher levels at  8, 20, 40, and 80 x
10$^{19}$ cm$^{-2}$, overlaid on an SDSS \textit{i}--band image.  The first contour also corresponds to a 8 $\sigma$ detection in three channels. The size of the D--array beam is indicated with the white circle.}
\label{arrow5} 
\end{minipage}
\vspace{0.45cm}
\begin{minipage}{\linewidth}
\centering
\includegraphics[scale=0.5] {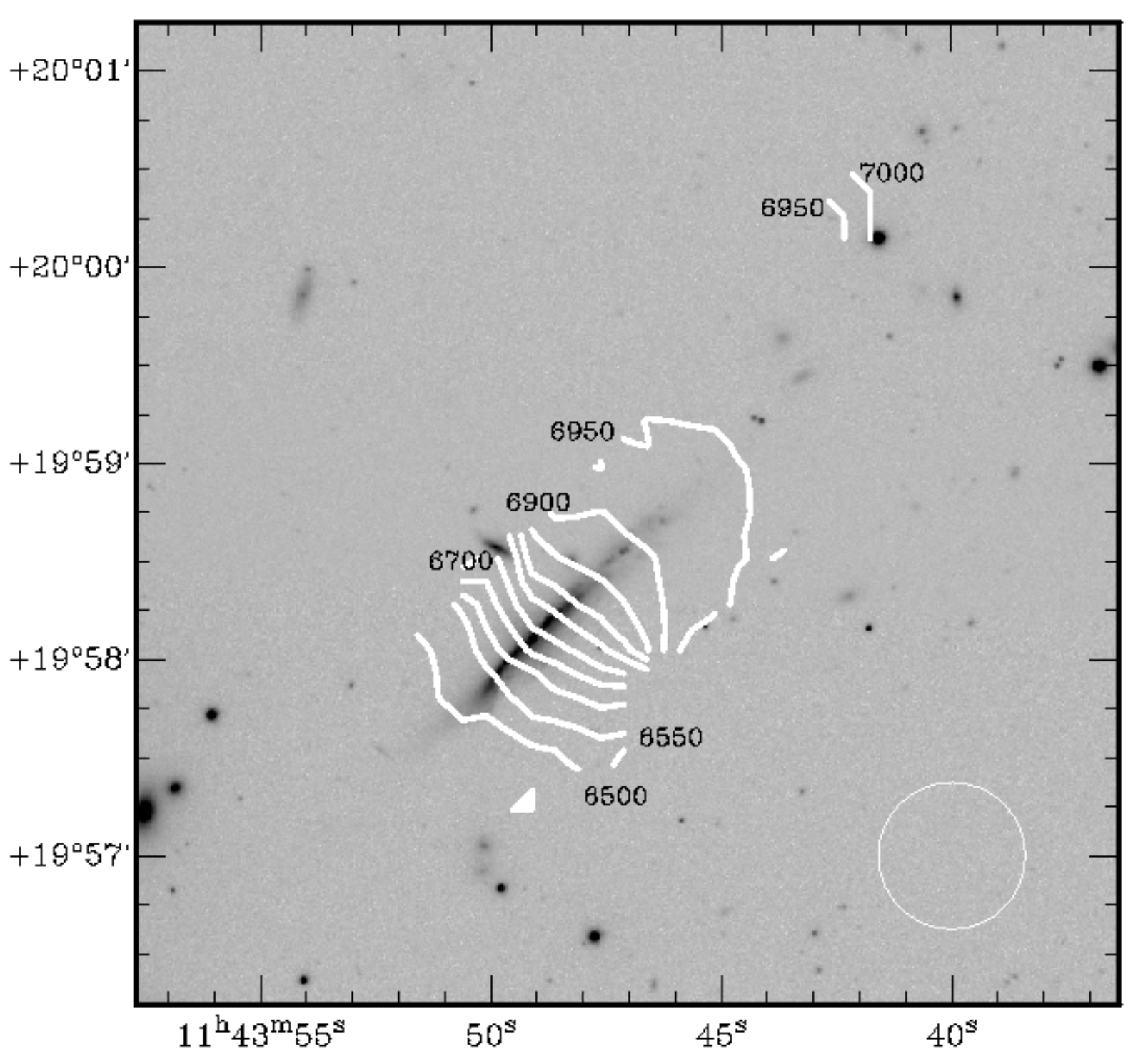}
\caption{\textbf{\cg087 velocity field:} White contours show velocities from 6500 to 7000 \km, in 50 \km\ steps based on the
robust 0 cube overlaid on an SDSS \textit{i}--band image. The size of the D--array beam is indicated with the white circle.}
\label{arrow35} 
\end{minipage}
\end{figure}

$\bullet$ \textbf{\cg087:} The integrated \hi\ map (Figure \ref{arrow5})
shows the \hi\ intensity maximum is offset by \aprox 30 arcsec (12 kpc) to the NW
of the nucleus, confirming the \hi\ is asymmetrically distributed, as earlier
reported by \cite{gava89} and \cite{dick91}. Higher spatial resultion and lower sensitivity \hi\ maps have previously been presented for the inner parts of this galaxy by \cite{hota07}. The main body of the galaxy displays rotation in
the velocity range \aprox 6500--6900 \km\ (Fig.\,\ref{arrow35}). We do not see the jump in velocities observed in \halpha\ and CO
\citep[Gavazzi \al 2001a;][]{bosel94}\nocite{gava2001a}, but this may be due to our
low spatial resolution. We are not able to confirm the merger hypothesis
suggested by the H$\alpha$ kinematics \citep[Gavazzi et al. 2001a \al;][]{amram02}\nocite{gava2001a}, but the \hi\
velocity field has a much steeper velocity gradient SE of the nucleus,
coinciding with the disturbed optical disk. We also see that the \hi\ disk is
sharply truncated, with its SE edge approximately coinciding with the edge of
the optical disk. An impressive \hi\ tail, coinciding with the radio continuum tail \citep{gava95}, extends at least 70 kpc to
the NW of the nucleus,  with an increasing offset from the plane of the
optical galaxy. At the end of this tail the \hi\ map shows a clump without an
optical counterpart. The clump has a
velocity width of \aprox 100\km. 

\begin{figure}
\centering
\vspace{12mm}
\includegraphics[scale=0.62] {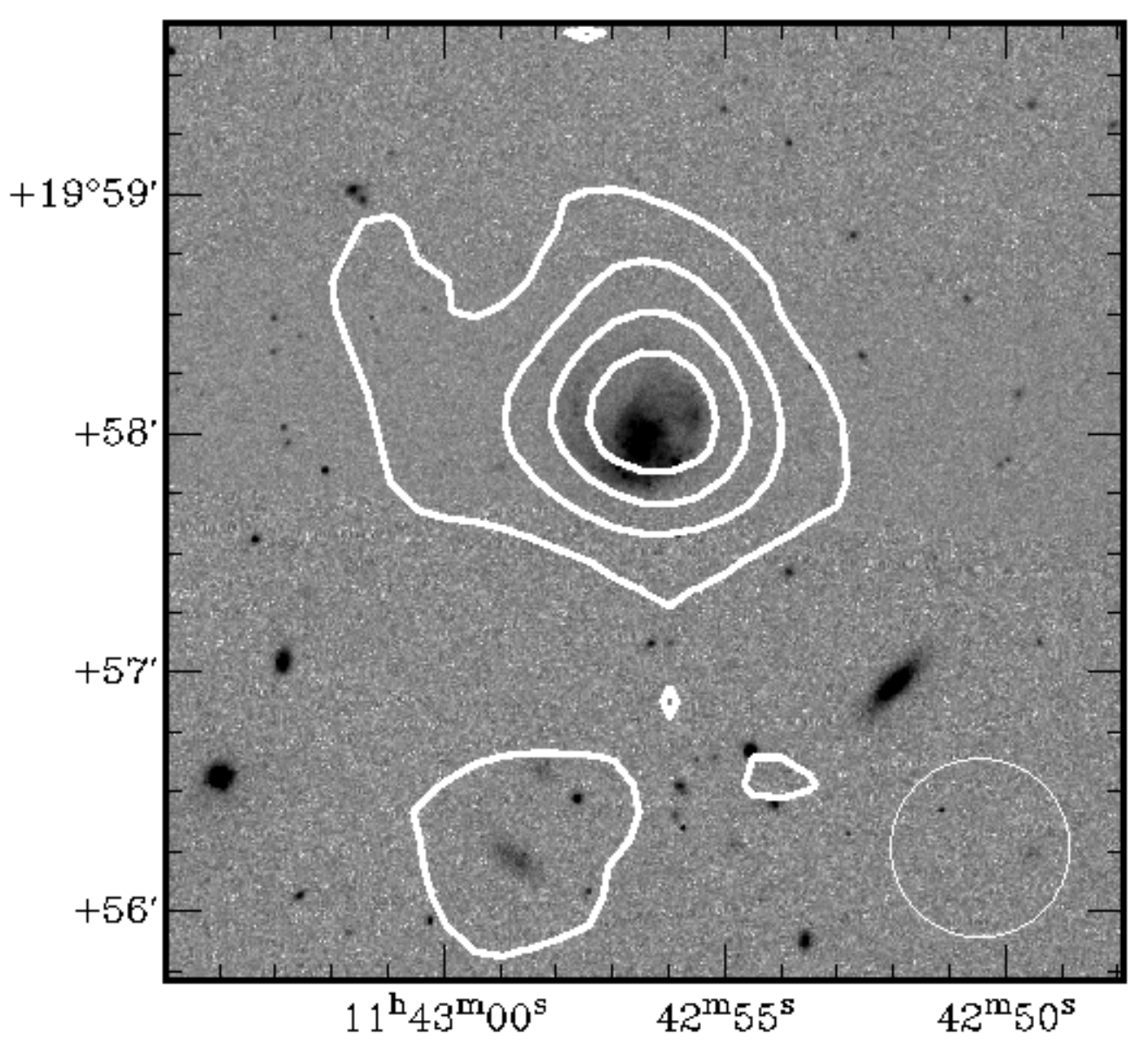}
\caption{\textbf{\cg073 (upper contour) and A\,1367\,[GP82]1292 (lower contour):} Contours trace the robust 0 \hi\ surface density, with the outer contour indicating a column density of  \textit{N}$_{HI}$  = 3 x 10$^{19}$ cm$^{-2}$, with higher
levels at 10, 17 and 24 x 10$^{19}$ cm$^{-2}$ overlaid on an SDSS
\textit{i}--band image.  The first contour also corresponds to a 7 $\sigma$ detection in three channels. The size of the D--array beam is indicated with the white circle. }
\label{73big} 
\end{figure}

$\bullet$ \textbf{\cg073 and A\,1367 [GP82]1292:} Figure \ref{73big} shows
the \hi\ in \cg073 having a slightly asymmetric distribution, the intensity
maximum being displaced \aprox 8 arcsec (3 kpc) to the north of the optical
disk, confirming the earlier report of \hi\ asymmetry \citep{dick91}. The \hi\ displacement is in the same direction as the \halpha\ and radio continuum tails (Gavazzi \al 1995, 2001b). \nocite{gava95,gava2001b}
\hi\ does not appear truncated, which is consistent with the normal \hi\
content (\defhi\ = 0.02). Also shown in the same figure is an \hi\ detection
at the position of the blue dwarf galaxy Abell [GP82]1292 which appears
undisturbed. Given the projected distance of \aprox 1.9 arcmin (47 kpc) and
the 11 \km\ velocity separation between the two (perturbation\footnote{ $p_{gg}= \frac{(M_{comp} / M_{gal}) }{(d/r_{gal} )^3} $ where $M_{gal}$ and $M_{comp}$ are the masses of the galaxy  and companion respectively, d is the separation and r is the galaxy disk radius \citep{byrd90}.}   parameter p$_{gg}$  $\sim$ 0.004), a gravitational interaction
cannot be ruled out.

\begin{figure}
\begin{minipage}{\linewidth}
\centering
\vspace{2.5mm}
\includegraphics[scale=0.4] {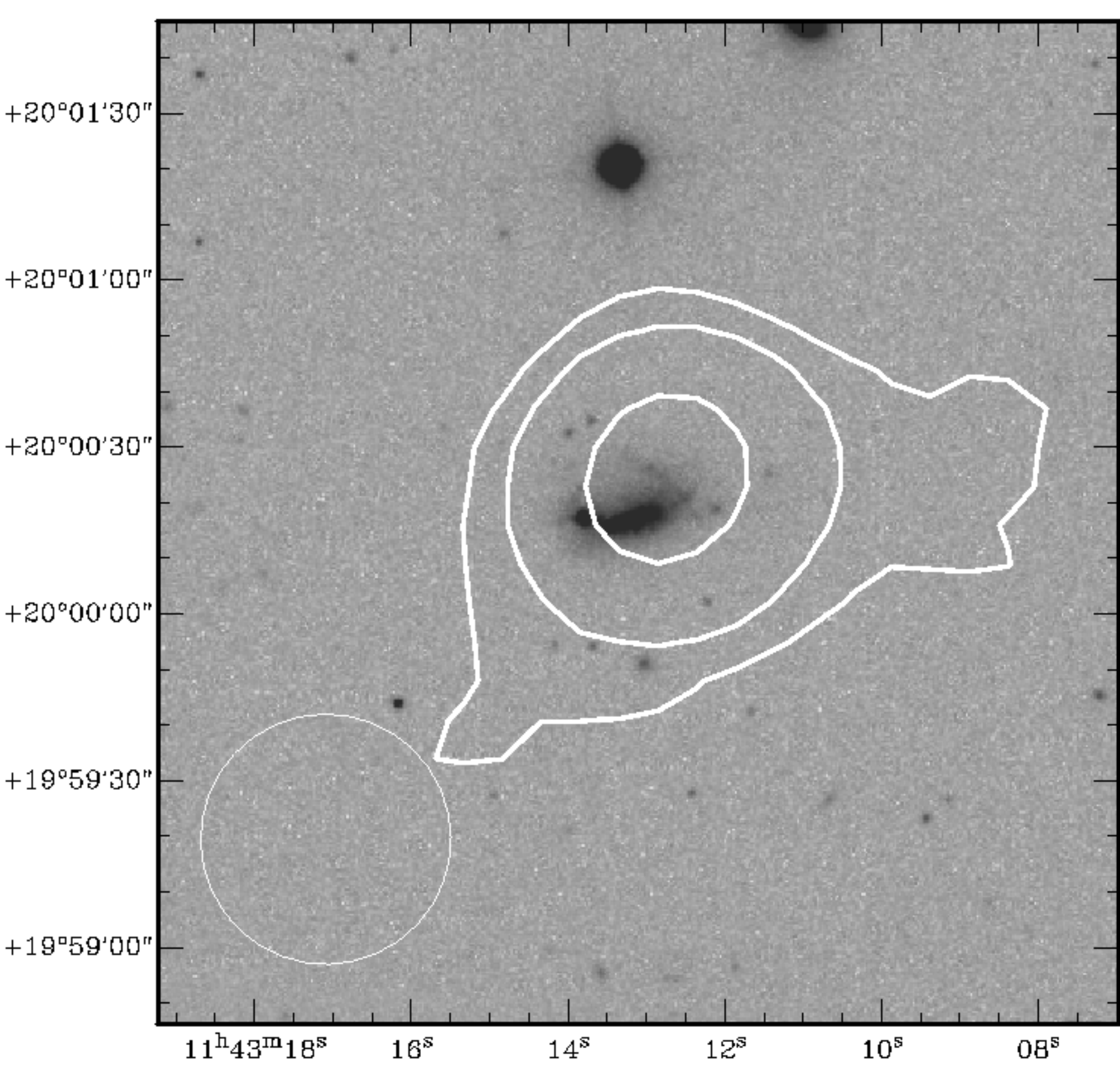}
\caption{\textbf{\cg079:} White contours are from a robust 0 \hi\ surface density map, with the outer contour indicating a column density of  \textit{N}$_{HI}$  = 3 x 10$^{19}$ atoms cm$^{-2}$, followed by contours at 10 and 17 x 10$^{19}$ atoms cm$^{-2}$ on an SDSS \textit{i}--band image.   The first contour also corresponds to a 7 $\sigma$ detection in three channels.   The size of the D--array beam is indicated with the white circle. }
\label{79big}
\end{minipage}
\end{figure}

$\bullet$ \textbf{\cg079:} \hi\ has been previously detected in this
galaxy \citep{gava89,dick91}. Figure \ref{79big} shows the \hi\ to be symmetrically
distributed but with the intensity maximum displaced by
\aprox 10 arcsec (4 kpc) to the NW of the optical disk, as previously reported by \cite{gava89} and \cite{hota07}. The \hi\ offset is in the same direction as the \halpha\ and radio continuum tails (Gavazzi \al 1995, 2001b) \nocite{gava95,gava2001b}.

\subsection{ \bf Giant late--type galaxies (\cg062, \cg068, \cg072 and
\cg091)} 

In addition to the three starbursts we were able to map four more spirals in
the NW and W sub--clusters (Figure \ref{bigxmm}).

\begin{figure}
\begin{minipage}{\linewidth}
\centering
\includegraphics[scale=0.4] {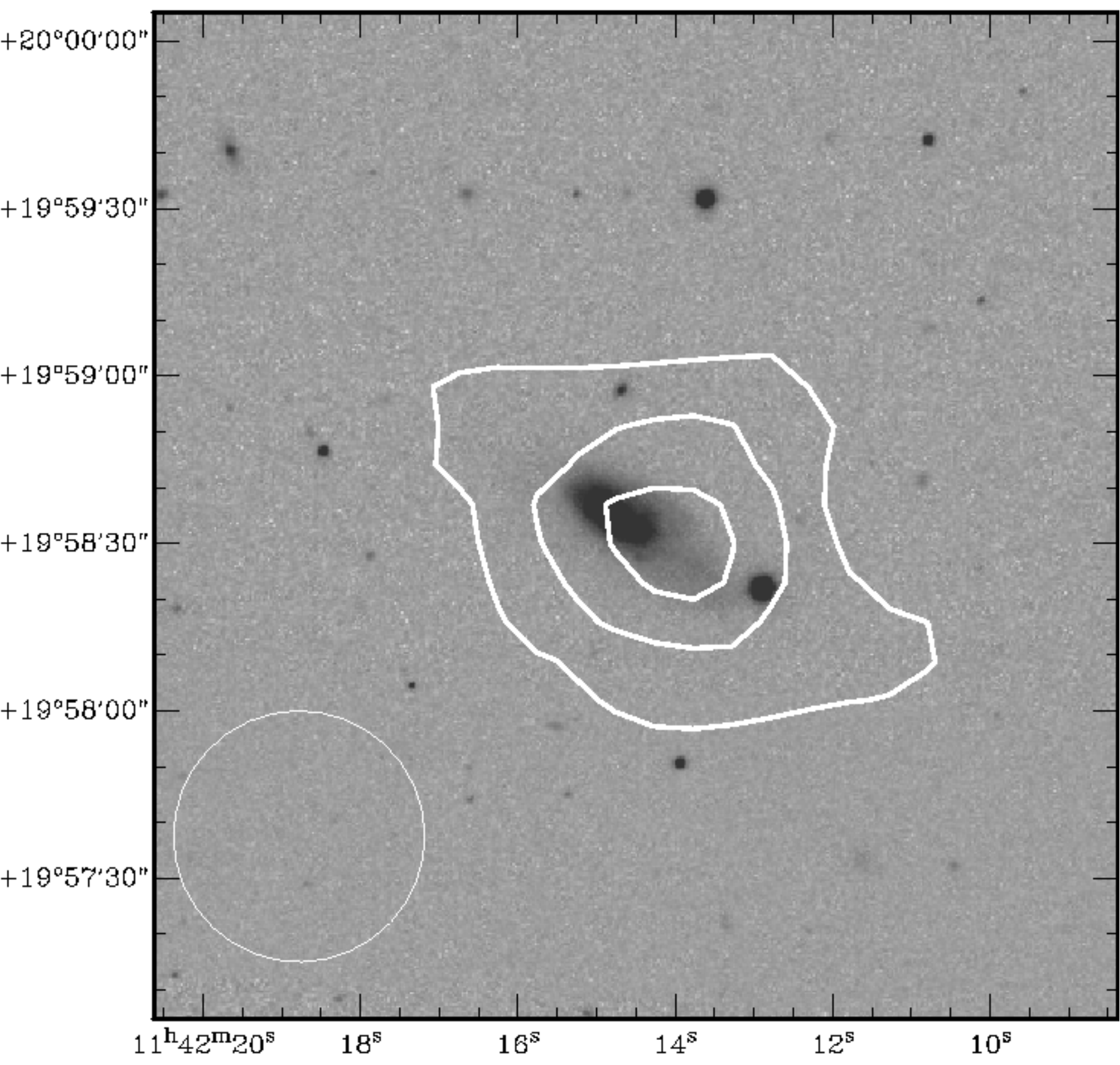}
\caption{{\bf \cg062:} White contours are robust 0 \hi\ surface
density levels with the outer contour corresponding to a column density of  \textit{N}$_{HI}$  = 3 x 10$^{19}$
atoms cm$^{-2}$, with higher levels at 8 and 13 x 10$^{19}$ atoms cm$^{-2}$,
on an SDSS \textit{i}--band image. The first contour also corresponds to a 2 $\sigma$ detection in three channels.  The size of the D--array beam is indicated
with the white circle.}
\label{62big}
\end{minipage}
\end{figure}

$\bullet$ \textbf{\cg062:} This optically asymmetric Sbc (NED) or Pec
Galaxy OnLine Database Milano Network (GOLDMine\footnote{http://goldmine.mib.infn.it/}), galaxy is well beyond Field A's FWHP beam, but our weak \hi\ detection is consistent with the \defhi\ (AGES) of 0.35. Figure \ref{62big} shows the
\hi\ intensity maximum displaced \aprox 15 arcsec (6 kpc) to the SW of its
optical counterpart, in the direction of the optical tail.  \cg062, like \cg087 has a pronounced asymmetric \hi\ spectrum \citep{gava89}.

\begin{figure}
\begin{minipage}{\linewidth}
\centering
\includegraphics[scale=0.4]{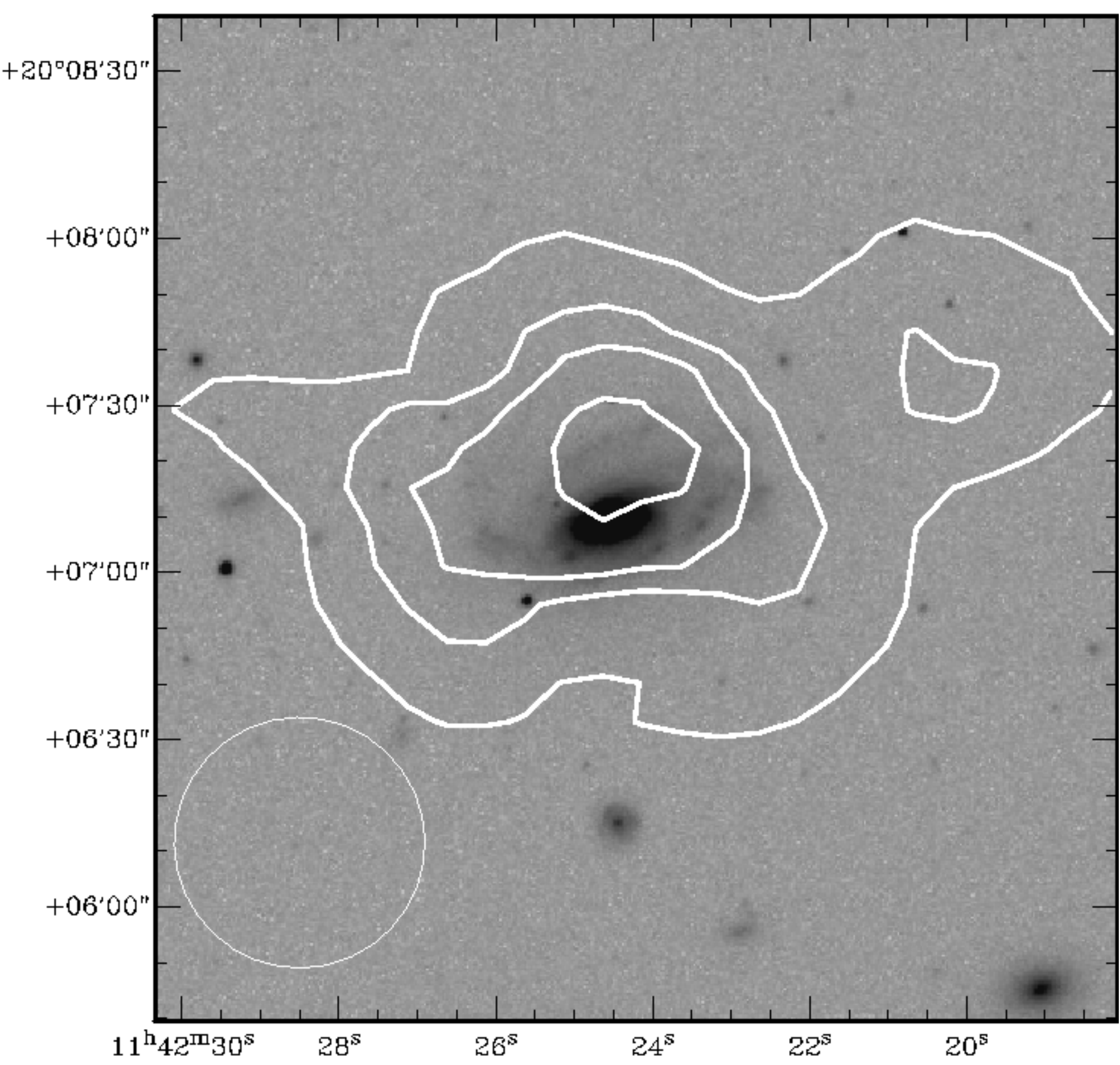}
\caption{\textbf{\cg068: } White contours are from a robust 0 \hi\
surface density map, with the outer contour indicating a column density of  \textit{N}$_{HI}$  = 10 x 10$^{19}$ atoms
cm$^{-2}$, higher levels at 30, 50 and 70 x 10$^{19}$ atoms cm$^{-2}$ on an SDSS
\textit{i}--band image.The first contour also corresponds to a 5 $\sigma$ detection in three channels.   The size of the D--array beam is indicated with the
white circle.}
\label{68big} 
\end{minipage}
\end{figure}

$\bullet$ \textbf{\cg068:} Despite its position well beyond the FWHP region of
the VLA's primary beam, \cg068's \hi\ signal was sufficient to show a double--peaked VLA spectrum which approximately matches its AGES counterpart,
strongly suggesting that this massive Sbc galaxy is inclined and  \hi\ rich
(\defhi\ of -0.28, Table \ref{sdss}). This galaxy was tentatively reported as
having its \hi\ intensity maximum displaced north of the optical disk
\citep{dick91}. The position of the VLA \hi\ intensity maximum \aprox 15 arcsec (6 kpc) in projection north of the optical nucleus confirms this (Figure \ref{68big}).

\begin{figure}
\begin{minipage}{\linewidth}
\centering
\includegraphics[scale=0.4]{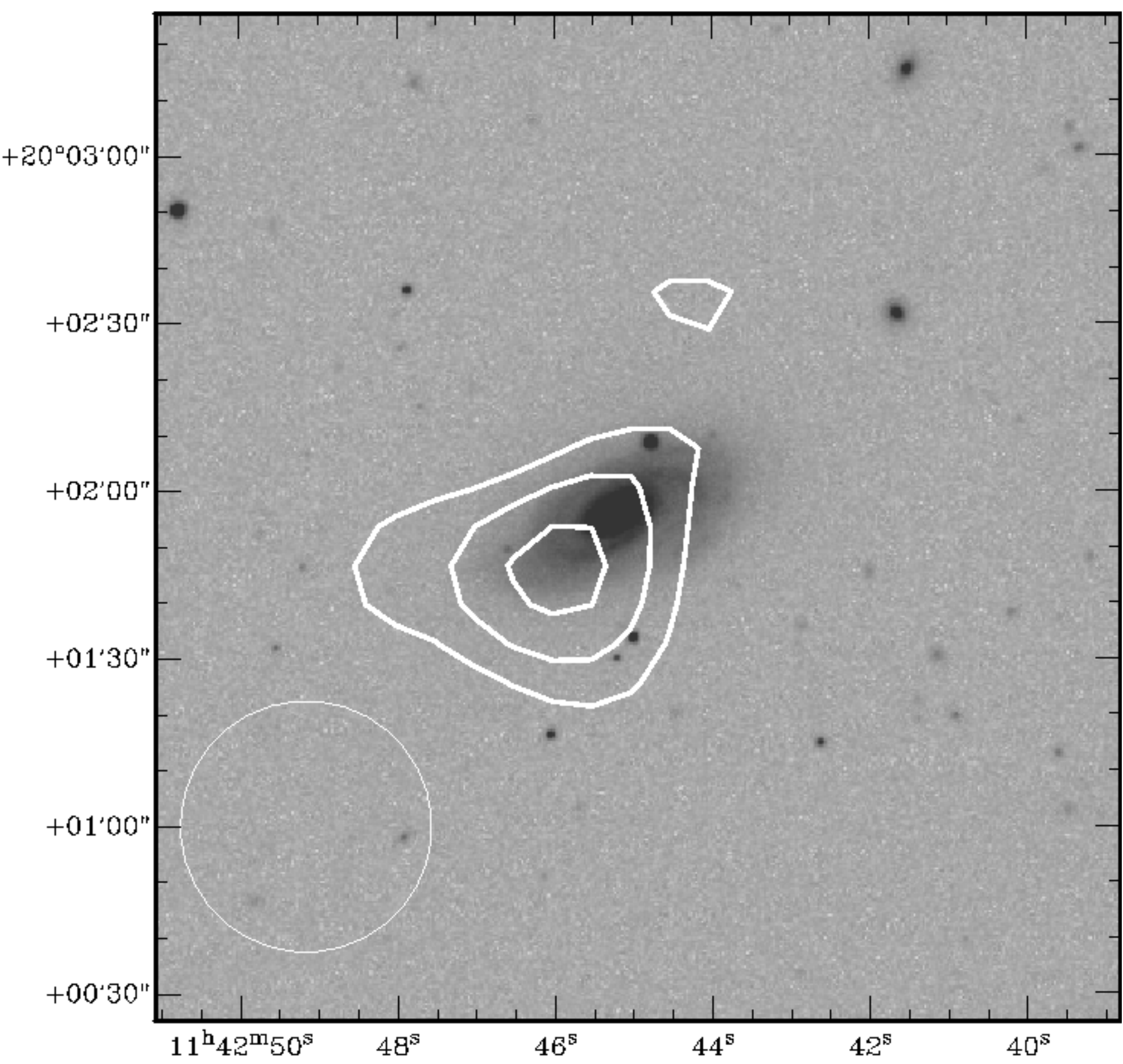} 
\caption{\textbf{ \cg072:} White contours are from a robust 0 \hi\
surface density map, with the outer contour indicating a column density of \textit{N}$_{HI}$  = 3 x 10$^{19}$
atoms cm$^{-2}$ and the higher contours are 5 and 7 x 10$^{19}$ atoms
cm$^{-2}$, on an SDSS \textit{i}--band image. The first contour also corresponds to a 4 $\sigma$ detection in three channels.    The size of the D--array beam is
indicated with the white circle.}
\label{72big}
\end{minipage}
\end{figure}

$\bullet$ \textbf{\cg072:} This Sa galaxy lies, in projection, close to
the W sub--cluster galaxy density maximum \citep{cort04}. Our \hi\ detection
is weak, consistent with its high \defhi\ of 0.55 (AGES). Figure \ref{72big}
shows an \hi\ intensity maximum offset \aprox 12 arcsec (5 kpc) to the SE of
the optical nucleus, and suggests a truncated \hi\ disk. However, this
requires further confirmation as the galaxy is beyond the  FWHP beam of Field A.
The high \defhi\ and anomalous direction of its \hi\ displacement relative to that seen in nearby spirals  is similar to NGC\,4848 in Coma where a truncated \hi\ disk was reported by \cite{bravo01} and further CO-imaging and numerical simulations by Vollmer \al (2001b) \nocite{voll01b} revealed a case of gas fallback.

\begin{figure}
\begin{center}
\includegraphics[scale=0.6] {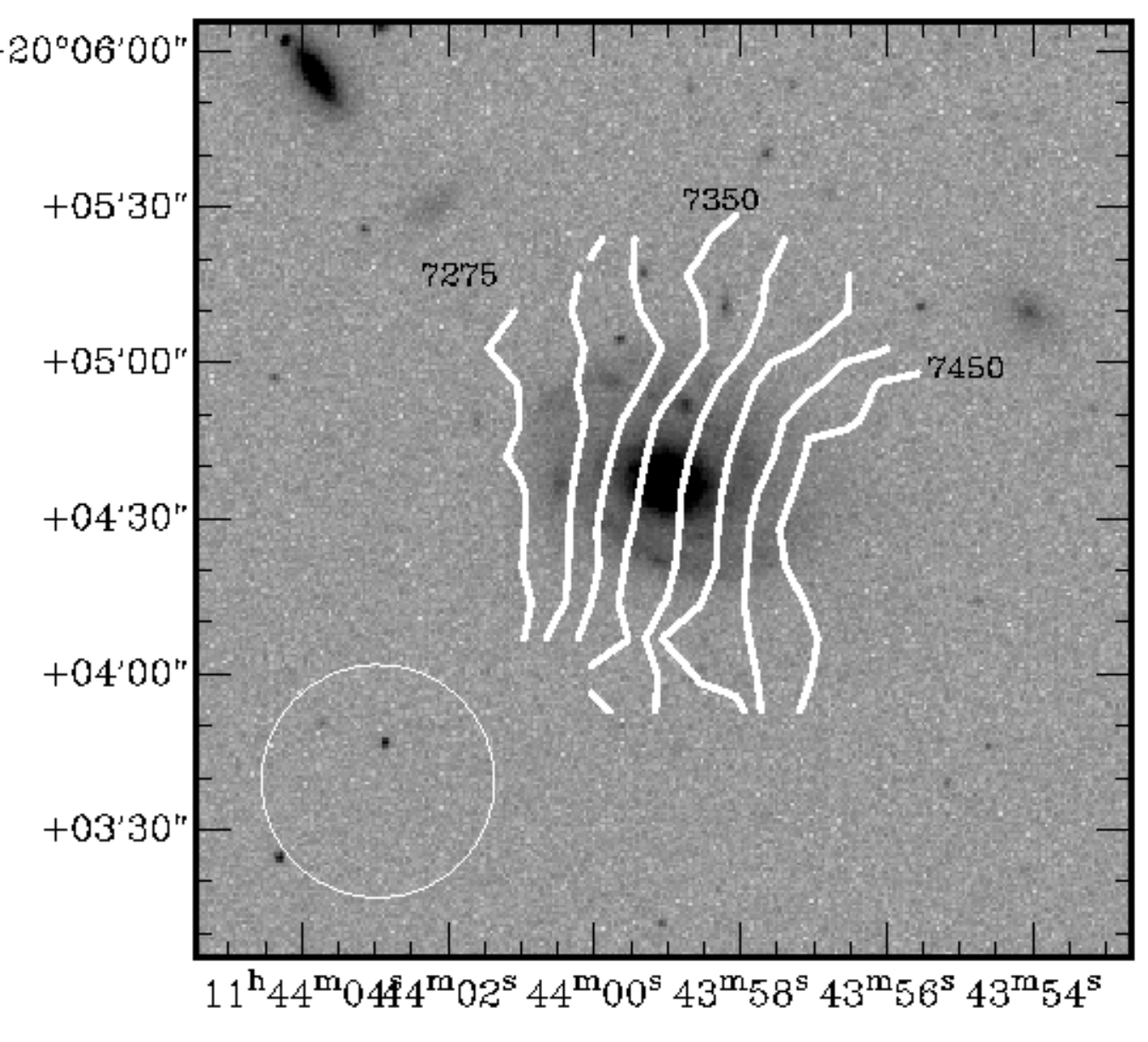}
\caption{{\bf \cg091} \hi\ velocity field: based on a robust 0 cube,
Contours are 7275, 7300, 7325, 7350, 7375, 7400, 7425 and 7450 \km\ on an
SDSS \textit{r}--band image. The size of the D--array beam is indicated with
the white circle.}
\label{97091vf}
\end{center}
\end{figure}

$\bullet$ \textbf{\cg091:} \hi\ has been previously detected in this galaxy
with Arecibo and the VLA \citep{dick91}, but was not  spatially resolved. The AGES \defhi\ of -0.23 shows the galaxy to be \hi\ rich. \cg091 has the most symmetric \hi\ morphology of our VLA detections, with a near perfect coincidence of position and velocity between the optical nucleus and the \hi\ intensity maximum, (Tables \ref{sdss} and \ref{table2}). These characteristics together with the velocity field (Figure \ref{97091vf}), which shows normal rotation,  combine to suggest \cg91 is a normal spiral which is not interacting with its environment in any significant way.

\subsection{ \bf The \hi\ around the Blue In-falling Group (BIG)} 
\label{bbig}
\begin{figure}
\begin{minipage}{\linewidth}
\centering
\includegraphics[scale=0.4] {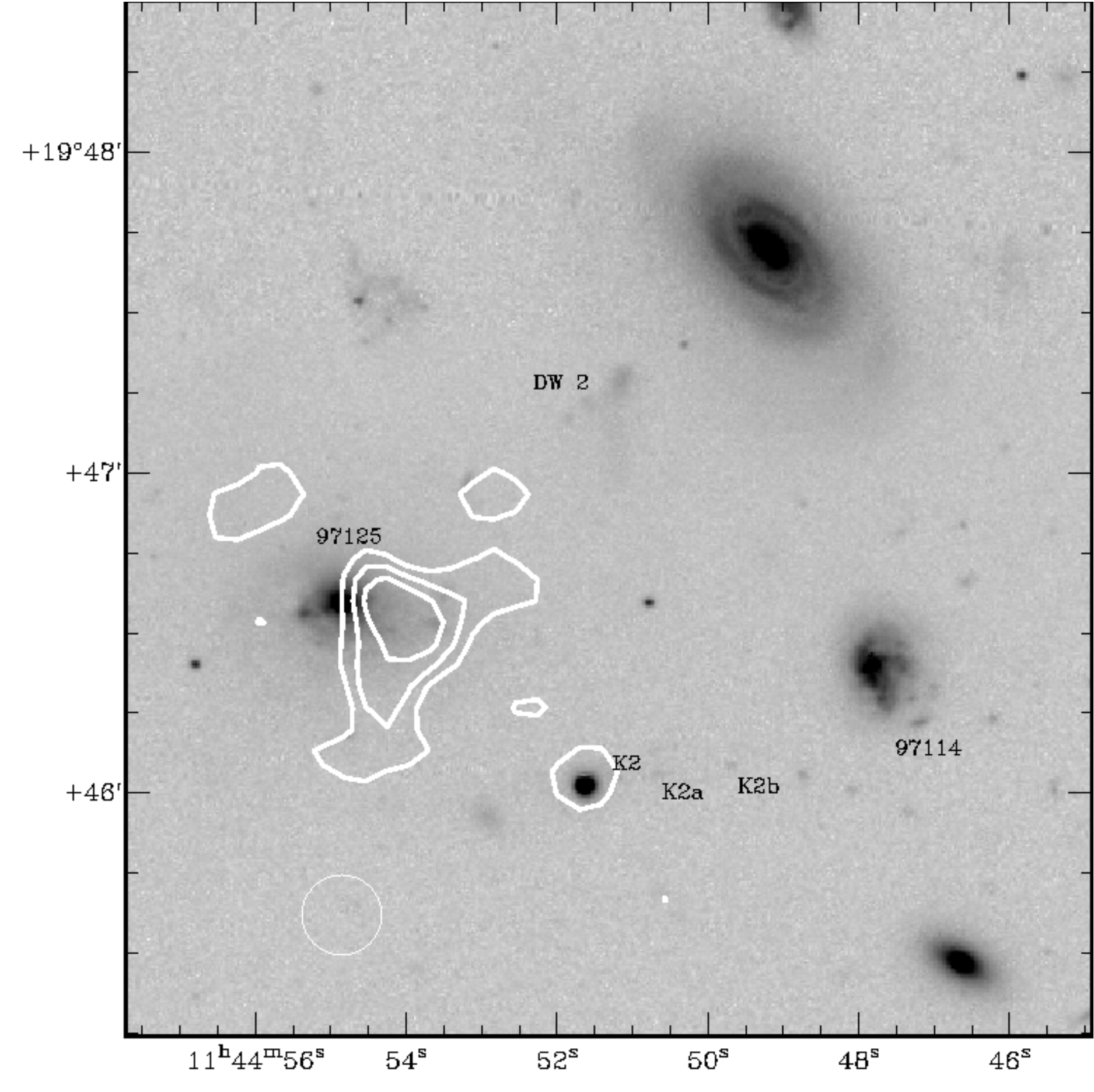}
\caption{{\bf \cg125 and knot near K2:} White contours are from a robust 0 \hi\ surface density map, with the outer contour indicating a column density of \textit{N}$_{HI}$  = 4 x 10$^{20}$ cm$^{-2}$, higher levels are at 8 and 12 x 10$^{20}$ cm$^{-2}$ on an SDSS \textit{g}--band image.  The first contour also corresponds to a 4 $\sigma$ detection in three channels.   The size of the C--array beam is indicated with the white circle.}
\label{big125}
\end{minipage}
\end{figure}

Our VLA \hi\ observation of the BIG (Field B) has a higher spatial resolution (\aprox 15 arcsec ), compared to Field A (\aprox 45 arcsec), but at significantly lower sensitivity because of the shorter integration time. The strongest detection in Field B is near the disturbed S0a galaxy
\cg125 and we confirm that the \hi\ intensity maximum is offset \aprox
12 arcsec (5 kpc) westward from the optical centre (Figure \ref{big125}). The high column density \hi\ to which Field B is sensitive is asymmetrically distributed with its major axis running approximately SE - NW. Comparing our \hi\ images to the earlier more sensitivity observations by \cite{sakai02} with the Westerbork Synthesis Radio Telescope (WSRT) indicates that BIG contains extensive diffuse \hi,  which our observation was not sensitive enough to detect. Comparison between our \hi\ spectrum and AGES confirms the presence of this diffuse \hi\ in the velocity range from 8000 to 8500 \km.
Further details of our observation of BIG can be found in Appendix A.

\subsection{ \bf Dwarf galaxy A\,1367\,[GP82]1227}
\label{dwf}
\begin{figure}
\begin{minipage}{\linewidth}
\centering
\includegraphics[scale=0.5] {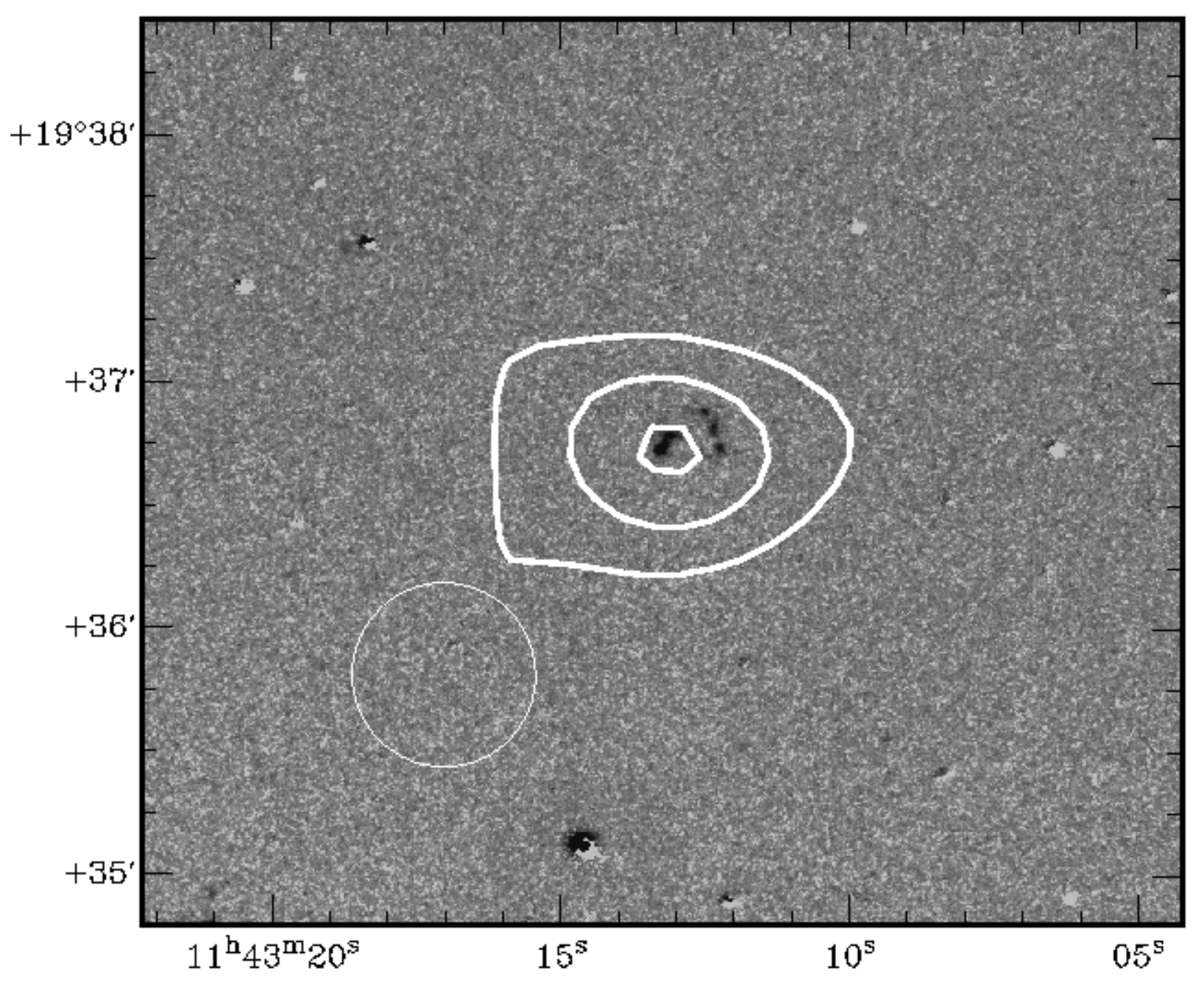}
\caption{\textbf{[GP82]1227:}White contours are from a robust 0 \hi\ surface density map. The outer contour is  \textit{N}$_{HI}$  = 3 x 10$^{19}$ cm$^{-2}$, higher levels are at 10 and 17 x 10$^{19}$ cm$^{-2}$, overlaid on an \halpha\ image. The first contour also corresponds to a 4 $\sigma$ detection in three channels.   The size of the D--array beam is indicated with the white circle.}
\label{1227small} 
\end{minipage}
\end{figure}

This isolated irregular dwarf galaxy is detected in \hi, both by the VLA and in
AGES, to the SW of the cluster centre in a region otherwise devoid of \hi\
detections (Figure \ref{1227small}). The VLA and AGES \hi\ spectra are in good
agreement and display a single narrow peak (W$_{50}$\,=\,56 $\pm$8 \km; Cortese \al 2008), typical of a dwarf irregular galaxy. The VLA observations show the \hi\ intensity maximum is offset 3 arcsec (1 kpc) to the south of its optical counterpart. Details of a low--resolution optical spectrum obtained for this galaxy can be found in Appendix B.

\newpage
\begin{figure*}
\begin{center}
 \includegraphics[ angle=0,scale=0.9] {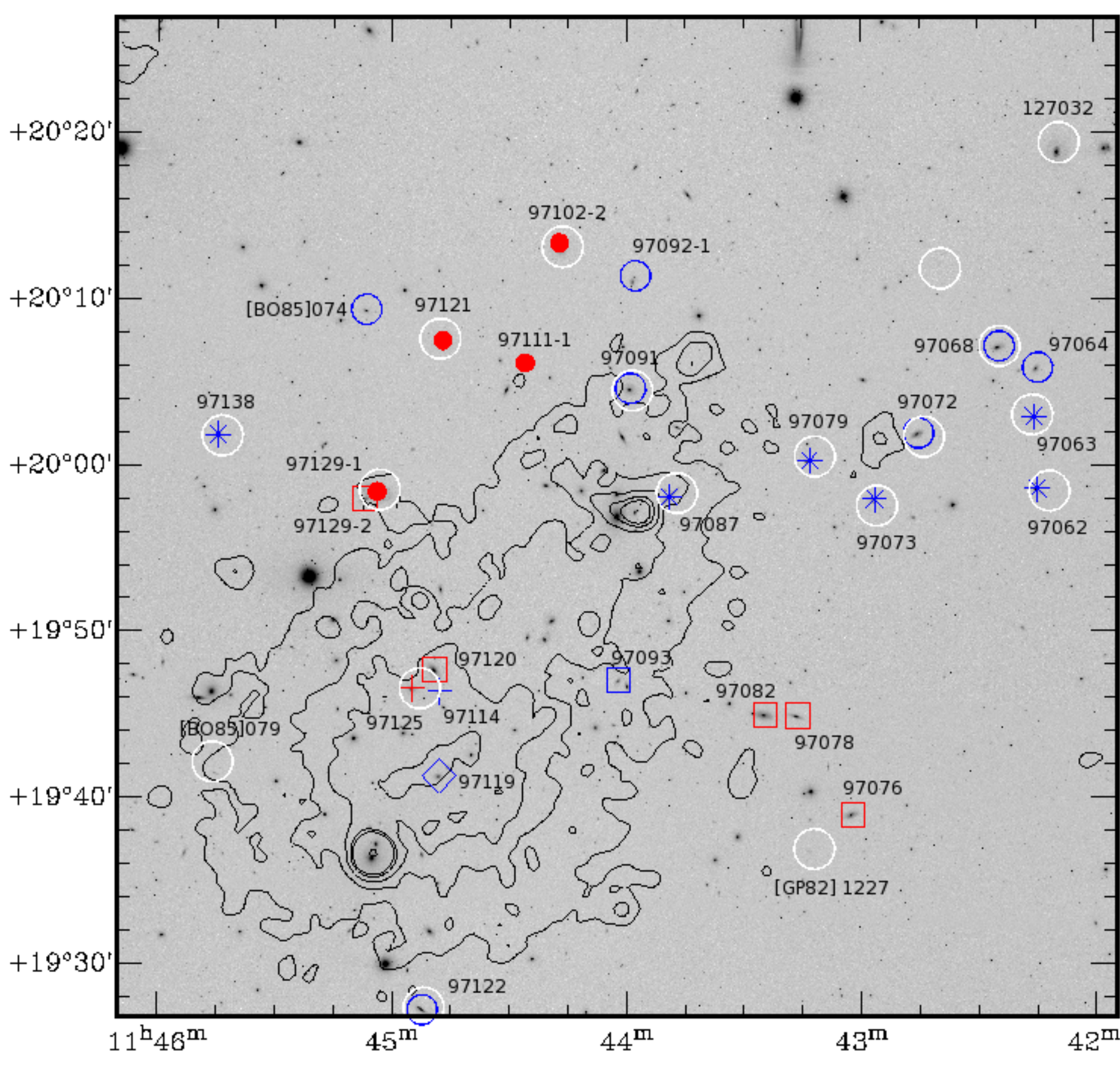}

\vspace{1cm}
\caption{Positions of selected late--type galaxies in
the volume sampled. The  \emph{colour} of the symbol (red or blue) used for each spiral indicates its SDSS \textit{g--i} colour: blue ($\leq$1.1), red ($>$1.1). AGES \hi\ detections are marked (white circles). The \emph{symbol} used for each spiral indicates its evolutionary state (discussed in section \ref{statesnew})  as follows; A--(asterisk), B--(open circle), C--(filled circle) and D--(open square)  BIG spirals--(cross) and unclassified--(diamond). } {\em ROSAT} X--ray intensities (black contours)
are overlaid on an SDSS \textit{i}--band image. Details for the galaxy
identifiers are given in note (a) of Table \ref{sdss}.
\label{groups}
\end{center}
\end{figure*}

\begin{table*}
\centering
\begin{minipage}{140mm}
\caption{ Parameters for the brightest late--type galaxies in A\,1367 }
\label{sdss}
\begin{tabular}{|l|l|l|r|r|l|l|l|r|r|lc} \hline
ID \footnote{Zwicky catalogue, except [BO85]074 from \cite{butch85}, with the suffixes indicating the NED pair identifier. } & ${\alpha}_{2000}$\footnote{Optical galaxy positions ex SDSS DR6} & ${\delta}_{2000}$ & Optical\footnote{from NED } & \textit{i}--band\footnote{ex SDSS DR6}&$g-i$\footnote{ex SDSS DR6}&Major\footnote{ex GOLDMine except \cg111 and [BO85]074 which are taken from NED} & Type\footnote{ex GOLDMine except [BO85]074 which is taken from NED}&H\,{\sc i} def \footnote{\defhi\ is calculated using the method of
\cite{hayn84} and the parameters from \cite{sola96}, using total \hi\ flux measurements from AGES and galaxy diameters equal to the major axis diameter from GOLDMine (except [BO85]074 which is taken from NED). For 97-079 the mass was
calculated from our VLA total flux because the AGES flux was affected by RFI. For
AGES non-detections, the \defhi\ lower limit was calculated using the AGES
detection limit of 6 x 10$^8$ \msolar. No \defhi\ could be calculated for \cg119 because flux from this galaxy was confused in the AGES beam with 3C264.  The calculated \hi\ deficiencies are highly uncertain  because of the intrinsic spread in \hi\ properties of individual spirals and we adopt an uncertainty of $\pm$ 0.3 dex throughout this paper.} & \multicolumn{2}{c}{Companion\footnote{A galaxy is considered to be a likely companion if its separation is $\leq$ 2 arcmin, the velocity difference is $\leq$ 500\km\ and the companion has a diameter $\geq$0.5 arcmin. }}&State\footnote{Evolutionary state described in section \ref{statesnew}  }\\ 
& & & velocity & [mag] & [mag] &
axis & & &\multicolumn{2}{c}{separation}&\\
&[$^h$ $^m$ $^s$] &[\degree\ \prim\ \prin\ ]&[\km ] &
&&[\prim]\ &&&[\km ]&[\prin\ ]& \\
\hline
97-064 &11 42 14.61 &20 05 52.18 &5976 &14.18 &1.06 &0.65 &S (dS) &$\geq$0.48&-&-&B\\
97-062 &11 42 14.78 &19 58 35.5 &7815 &14.78 &0.62 &1.01 &Pec &0.35&-&-&A\\
97-063 &11 42 15.62 &20 02 55.24 &6102 &15.27 &0.63 &0.58 &Pec &0.01&-&-&A\\
97-068 &11 42 24.49 &20 07 09.55 &5974 &13.24 &0.99 &1.23 &Sbc &-0.28&-&-&B\\
97-072 &11 42 45.16 &20 01 56.35 &6332 &13.61 &1.07 &1.21 &Sa &0.55&-&-&B\\
97-073 &11 42 56.45 &19 57 58.39 &7275 &15.08 & 0.55 &0.76 &Pec &0.02&-&-&A\\
97-076 &11 43 02.14 &19 38 59.02 &6987 &13.56 &1.17 &1.20 &Sb &$\geq$0.88&-&-&D\\
97-079 &11 43 13.08 &20 00 17.46 &7000 &14.37 & 0.47 &0.75 &Pec &0.25&-&-&A\\
97-078 &11 43 16.22 &19 44 55.46 &7542 &13.55 &1.21 &1.88 &Sa &$\geq$0.96&-&-&D\\
97-082 &11 43 24.55 &19 44 59.35 &6100 &13.00 &1.20 &1.30 &Sa &$\geq$0.77&-&-&D\\
97-087 &11 43 49.07 &19 58 06.49 &6725 &13.19 &0.64 &2.00 &Pec &0.39&-&-&A\\
97-092-NED1 &11 43 58.2 &20 11 08 &6487 &14.95 &0.99 &0.76 &Sbc &$\geq$0.62&-&-&B\\
97-091 &11 43 58.96 &20 04 37.34 &7368 &13.26 &0.99 &1.12 &Sa &-0.23&-&-&B\\
97-093 &11 44 01.95 &19 47 03.94 &4909 &15.05 & 0.42 &0.96 &Pec &$\geq$0.78&-&-&D\\
97-102-NED2 &11 44 16.49 &20 13 00.73 &6364 &13.67 &1.24 &1.08 &Sa &0.41&4&24&C\\
97-111 NED1 &11 44 25.93 &20 06 09.64 &7436 &14.92 &1.14 &0.45 &Irr& $\geq$0.21&194&24&C\\
97-121 &11 44 47.04 &20 07 30.32 &6571 &12.86 &1.18 &1.20 &Sab &0.38&-&-&C\\
97-114 &11 44 47.8 &19 46 24.31 &8293 &15.12 & 0.57 &0.54 &Pec &$\geq$0.34&22&102&-\\
97-119 &11 44 47.95 &19 41 18.56 &4895 &14.18 &0.98 &0.60 &Sa &$\geq$0.37&-&-&-\\
97-120 &11 44 49.16 &19 47 42.14 &5595 &12.53 &1.12 &1.32 &Sa &$\geq$0.78&-&-&D\\
97-122 &11 44 52.23 &19 27 15.12 &5468 &13.65 &0.93 &1.45 &Pec &0.47&-&-&B\\
97-125 &11 44 54.85 &19 46 35.18 &8271 &13.81 &1.20 &0.84 &S0a &$\geq$-0.14&22&102&-\\
97-129-NED1 &11 45 03.88 &19 58 25.24 &5085 &12.29 &1.21 &2.36 &Sb &-0.07&-&-&C\\
97-138 &11 45 44.6 &20 01 50.8 &5313 &13.99 & 0.42 &0.75 &Pec &-0.12&-&-&A\\
BO85 074 &11 45 06.57 &20 09 23.21 &6151 &14.69 &1.03 &0.38 &S &0.08&-&-&B\\
97-129 NED2 &11 45 06.95 &19 58 0.66 &7546 &14.66 &1.12 &1.07 &Sbc &$\geq$0.8&-&-&D\\
\\\hline
\end{tabular}
\end{minipage}
\end{table*}

\section{\bf \hi\ (AGES) and colour distribution in A\,1367} 
\label{hiages}

To place our VLA observations in context we used the AGES \hi\ survey together with SDSS data to analyse the \hi\ content and colour of bright late--type galaxies throughout the central volume of the cluster (including the VLA fields).\\
As part of the AGES project the 305-m Arecibo Telescope was used to survey \hi\
in a volume centred on A\,1367 covering about 5 square degrees, between
11$^h$34$^m$00$^s$\,$< \alpha_{2000}$\,$<$\,11$^h$54$^m$15$^s$, 19$^d$15$^m$\,$<\delta_{2000}$\,$<$20$^d$20$^m$, in a velocity range \aprox 1100--19000 \km\
\citep{cort08}. The survey has an angular resolution of 3.3\,$\times$\,3.8
arcmin and a velocity resolution of \aprox 10\,\km\ with a mass sensitivity limit of  $6 \times 10^8$ \msolar\ of \hi\ over a velocity W$_{50}$ = 200 \km. Positional uncertainties for the detections are estimated to be \aprox 18 arcsec (\aprox
7 kpc). The observations were made in drift scan mode providing a
consistent flux limited \hi\ detection threshold throughout the AGES volume. The observations and data reduction
process is described in \cite{auld06}. More details on this survey are given
in \citet{cort08}.

In the field, significant amounts of \hi\ are expected to be present in the disks of both spiral and dwarf galaxies. In the case of spirals the amount of \hi\ can be approximately related to the optical disk diameter and the galaxy's
morphological type \citep{hayn84,sola96}, but for dwarfs the
\hi\ mass fraction is both larger and more uncertain. The AGES detection
limit implies that spiral galaxies in A\,1367 with an optical disk diameter $<$
0.37 arcmin (\aprox 9 kpc) would not be detected in AGES \citep{sola96,cort08}. Galaxies with disk diameters below this limit are
almost certainly dwarfs, so that AGES is essentially only sensitive to \hi\ in late--type spiral
galaxies of A\,1367. However two dwarf galaxies were detected within our sampled
volume, showing that AGES includes galaxies from the high end of the dwarf \hi\ mass function.

AGES allows us to investigate the \hi\ deficiency of A\,1367's spirals, down
to its sensitivity limit, by comparing their expected \hi\ content with
actual AGES determined \hi\ content. To do this a catalogue of A\,1367's brightest spiral members was compiled, with the following criteria. We intially selected late--type galaxies with an SDSS $g$--band magnitude $<$ 15.5 within the sampled volume,
i.e., 1 square degree surrounding the NW sub--cluster core in the velocity
range 4000--9000\,\km, (approximately six times the velocity dispersion of
the cluster). Only those objects with an optical disk diameter $>$ 0.37 arcsec from GOLDMine (Gavazzi \al 2003b) \nocite{gava03b} or, if
unavailable, from the NASA Extragalactic Database (NED), were selected. For the selected late--type galaxies we determined their \defhi\ by comparing expected and observed \hi\ content
using the method of \cite{hayn84}. Hubble type was taken from
GOLDMine or, if unavailable, from NED. We selected all galaxies with Hubble type
later than S0. The only exception, \cg125, an S0a, was included
because of its unusually large \hi\ mass.

Applying these criteria we found the 26 bright late--type galaxies listed in Table \ref{sdss} which gives details of their SDSS
\textit{g} band magnitude and $g$--$i$ colour, optical diameter from GOLDMine/NED,
\hi--mass based on AGES flux, \defhi\ (or lower limit for AGES
non--detections) based on AGES flux calculated using the method from
\cite{hayn84} and parameters from \cite{sola96}, and separation parameters
for galaxies with a close companion.  A $g$--$i$ colour of 1.1 approximately coincides with the red sequence threshold displayed in the colour-magnitude plot in Figure 12 of \cite{cort08}. For our sample, those galaxies with $g$--$i$ colour $<
1.1 $ are referred to throughout the paper as {\it blue galaxies} and the rest
referred to as {\it red galaxies}. 

Figure \ref{groups} shows the  optical positions of the 26 spirals (red and blue symbols) and the positions of the 18 AGES \hi\ detections (white circles). In the figure the positions of the 16 selected galaxies with blue SDSS \textit{g--i} colours are indicated with blue symbols and the 10 with red SDSS \textit{g--i} colours with red symbols. Figure \ref{groups} also indicates the distribution of ICM
gas in the cluster based on \textit{ROSAT} X--ray observations (black contours).

A shift in position in Fig. \ref{groups} between an AGES \hi\ detection (white circle) relative to
its respective optical position (red and blue symbols) is an indication of a real displacement when
the offset exceeds 18 arcsec (7 kpc), the uncertainty in the AGES
positions. This is only the case for \cg087 and CGCG 127--032. However the much higher pointing accuracy of the VLA, \aprox 4 arcsec (1.6 kpc), reveals that \cg062, \cg068, \cg072, \cg073, \cg079, \cg087  and \cg125 all have \hi\ intensity maxima offsets relative to their optical counterparts larger than 3 $\sigma$. In all cases the offset direction is consistent with that from the higher uncertainty AGES data. The projected magnitude and direction of the VLA offsets given in Table \ref{table2} is consistent with that expected for spirals on radial orbits experiencing ram pressure stripping, except \cg062, \cg068, \cg072 where the offset direction is not radially with respect to the cluster centre.

The four AGES detections in Figure \ref{groups} without a selected late--type galaxy counterpart are the S0 galaxy CGCG 127-032, the two dwarfs, [GP82] 1227 and [BO85] 079, and an AGES detection to the NE of \cg068. \cg119 was not detected in the AGES survey because of its proximity to 3C264 and is therefore not included in our analysis.\\

\section{Discussion}
\label{discu}
The projected distribution, colour, \hi\ content and velocities of the selected spirals
throughout the sampled volume on large and small scales provides information
on the recent history of the cluster as well as clues to the
mechanisms impacting the galaxies' ISM. In the following subsections we
consider the global \hi\ properties, the role of ram pressure stripping attributable to the NW subcluster ICM and the evolutionary state of the spirals.\\

\subsection{Global \hi\ properties }
\label{golb}

Considered on a scale of several megaparsecs the pattern of \hi\ deficiencies projected in RA for Coma and A\,1367 in Figure 2 of \cite{gava89} show striking differences. From their figure it is clear that with few exceptions the Coma spirals across the central \aprox 5\degree\ of cluster are significantly \hi\ deficient ($>$ 0.3). In contrast for A\,1367 the significantly \hi\ deficient spirals are found in a more narrow range of projected RA (\aprox 1\degree) and even at its centre a majority of the spirals are  not deficient (\hi\ deficiencies $<$ 0.3)

Within our sampled volume for A\,1367 (\aprox\ central 1.5 Mpc) the bright spirals, irrespective of \hi\
content, are found preferentially projected onto the northern half  (Figure \ref{groups}), with the AGES detections (white circles in Figure \ref{groups}) highly concentrated in a band running
roughly E--W between declinations of 19$^\circ$55\prim\ to
20$^\circ$15\prim.\\

\begin{figure}
\begin{center}
\includegraphics[ scale=0.35] {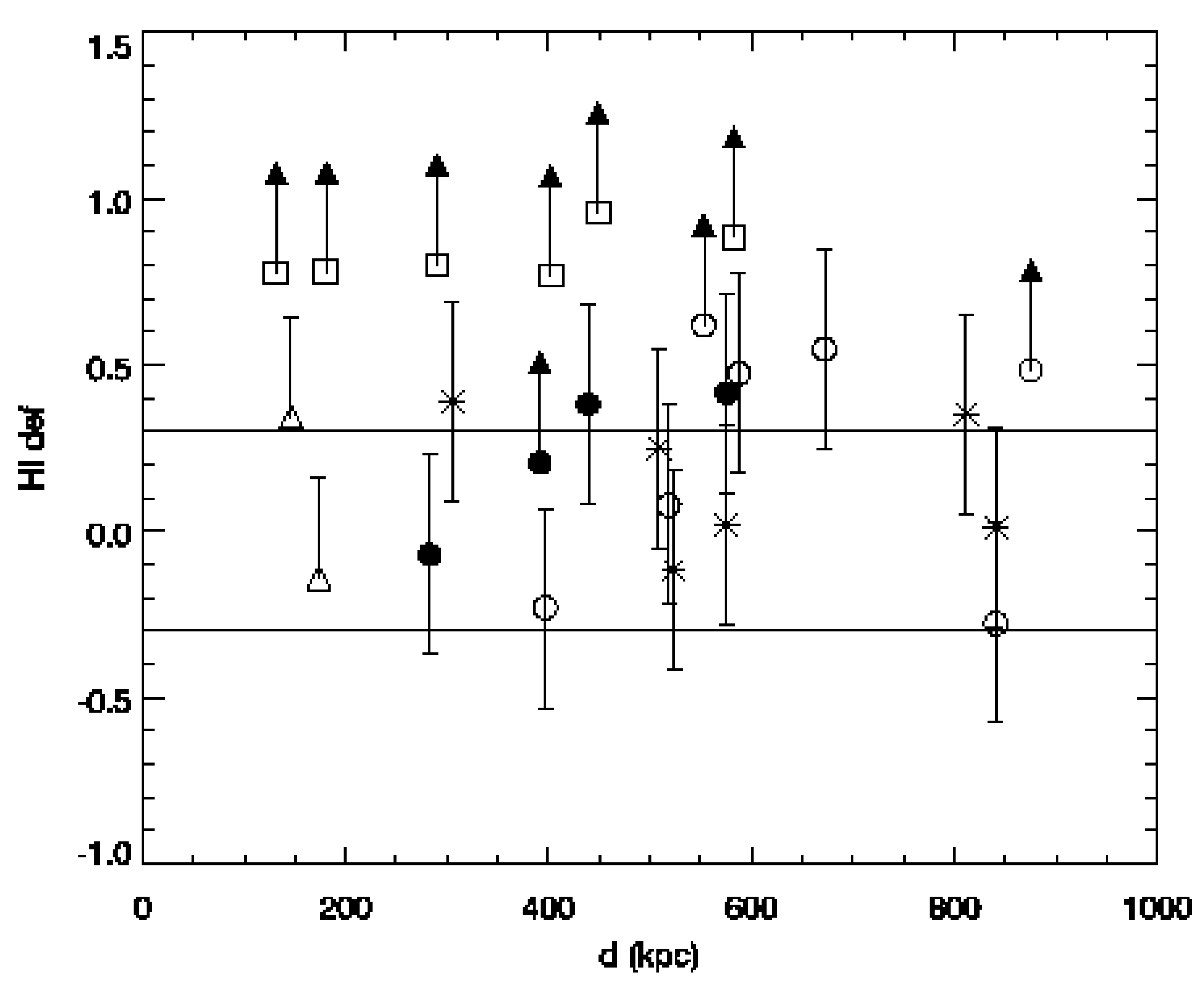}
\caption{The \defhi\ of all the selected late-type galaxies in A 1367, (excluding \cg119) as a function of projected distance from the cluster centre. The following symbols are used to indicate the state (discussed in section \ref{statesnew}) of each spiral: A--(asterisk), B--(open circle), C--(filled circle) and D--(open square). For AGES non-detections the \defhi\ is a lower limit and is indicated with   arrowheads. The two BIG spirals are confused in the AGES beam (triangles with upper error bar only) and in both cases the value is the lower limit for
\defhi. The region between the horizontal lines approximates the natural range of variation expected in the field population.}
\label{center_def}
\end{center}
\end{figure}

As an initial step to understand the global \hi\ properties in the sampled volume we
plotted, in Figure \ref{center_def}, the \defhi\ of all the selected
late--type galaxies as a function of projected distance from the cluster centre, with
lower limits for \hi\ non--detections, and BIG members confused in the AGES
beam. The plot shows that the mean \defhi\ of the selected galaxies is
appreciably greater than in the field indicating that the cluster environment
has impacted the ISM of a significant fraction of them.  However, unlike Coma where a clear increase in \defhi\ is seen within a radius of $\sim$1 Mpc of the core \citep{bravo00}, there is no obvious correlation 
between \defhi\ and projected proximity to the cluster core. It should be noted, though, that  the two interacting major sub--clusters revealed by the X--ray observations (see Fig. \ref{bigrosat}) complicate the
investigation of any correlation between the cluster's ICM density and
\defhi\ and in section \ref{rpsnw} we try to address the effects which are specific to the NW sub--cluster.

\subsection{Ram pressure stripping by the NW subcluster ICM }
\label{rpsnw}
We are interested to know if the observed \hi\ deficiencies can be explained in terms of ram pressure stripping. In
order to quantify the role of this mechanism around the NW sub--cluster, we
carried out a simple estimate of the strength of ram pressure as a function
of radius from the NW sub--cluster ICM core. The ICM density
($\rho_\mathrm{ICM}$) distribution was estimated using the
hydrostatic-isothermal $\beta$-model of \cite{cava76} with the gas profile
parameters from \cite{don98}. Ram pressure values were derived from Gunn \&
Gott's (1972) equation: $ P_{ram} = \rho_{ICM} \times v_{rel}^2$, where
$\rho_{ICM}$ is the density of the ICM at the galaxy position and $v_{rel}$ 
is the galaxy's velocity relative to the ICM. We have no information about
$v_{rel}$ for the selected galaxies, although we consider a
galaxy's radial velocity, relative to the cluster systemic velocity, as a
lower limit for $v_{rel}$.

\begin{figure}
\centering
\includegraphics[scale=0.7] {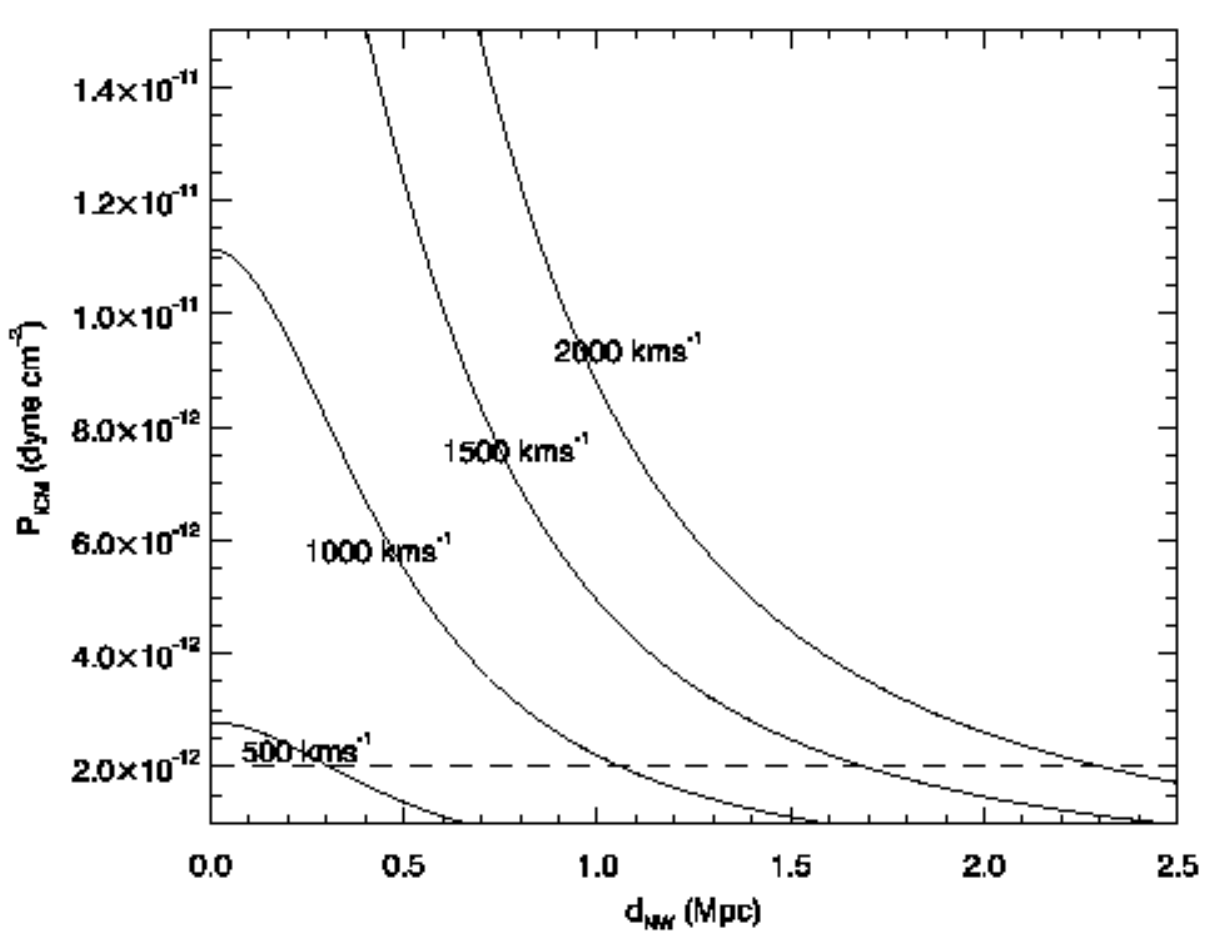}
\caption{Ram pressure as a function
of the distance from the X--ray emission maximum of the NW sub--cluster for
$v_{rel}$ of 500, 1000, 1500 and 2000\,\km. The dashed line indicates the threshold for ram pressure stripping based on \citet{caya94}. } 
\label{pram} 
\end{figure}

The results of this modelling of $ P_{ram}$ as a function of proximity to the
NW core for four different values of $v_{rel}$, up to 2000\,\km\
(approximately the radial velocity difference of the BIG), are shown in
Figure \ref{pram}. A spiral near the NW sub--cluster is expected
to suffer ram pressure stripping if $ P_{ram}$ values exceed
the restoring force of the galaxy's stellar and gas disks, at the gas disk's periphery, defined as
$F_r\,=\,2\pi\,G\,\sigma_{stellar}\,\sigma_{gas}.$ The threshold ram pressure required to remove \hi\ from the outer edge of the disk can be estimated from the  typical
value for the restoring force per unit area of $F_r\,=\,2\,\times\,10^{-12}$\,dyn\,cm$^{-2}$, obtained by \citet{caya94} for spiral galaxies in Virgo (dashed line in Figure \ref{pram}). The crossing time for the cluster is 1.7 Gyr (Boselli $\&$ Gavazzii 2006a) \nocite{bosel06a} making it likely that spirals within the central Mpc have been subject to ram pressure over periods of several times  $10^{8}$ yrs.  Modelling  by \cite{roed07} indicates that for a mid--size spiral ($V_{rot}$ = 200\km ) sustained exposure to ram pressures over periods of the order of $10^{8}$ yrs of $10^{-12}$\,dyn\,cm$^{-2}$ will result in only minor stripping of \hi, pressures of $10^{-11}$\,dyn\, cm$^{-2}$ will cause significant stripping, while $10^{-10}$\,dyn\, cm$^{-2}$ will almost completely remove the \hi. Massive galaxies are expected to suffer proportionately less \hi\ removal, and less massive galaxies proportionately more.

The implication of Figure \ref{pram} and the \cite{roed07} models is that any galaxy with $v_{rel}$ $\geq$ 1000 \km\ anywhere within a radius of 1 Mpc of the NW sub--cluster will undergo weak to moderate ram pressure stripping. But removal of large fractions of \hi, i.e.  $ P_{ram}$ $\geq$ $10^{-11}$\,dyn\,cm$^{-2}$, would require a transit of the high density ICM within \aprox\ 0.75 Mpc of the cluster centre  at a velocity in excess of 1500\km\ or transit within \aprox\ 0.25 Mpc of the core at 1000 \km.

\begin{figure}
\begin{center}
\includegraphics[ scale=0.35] {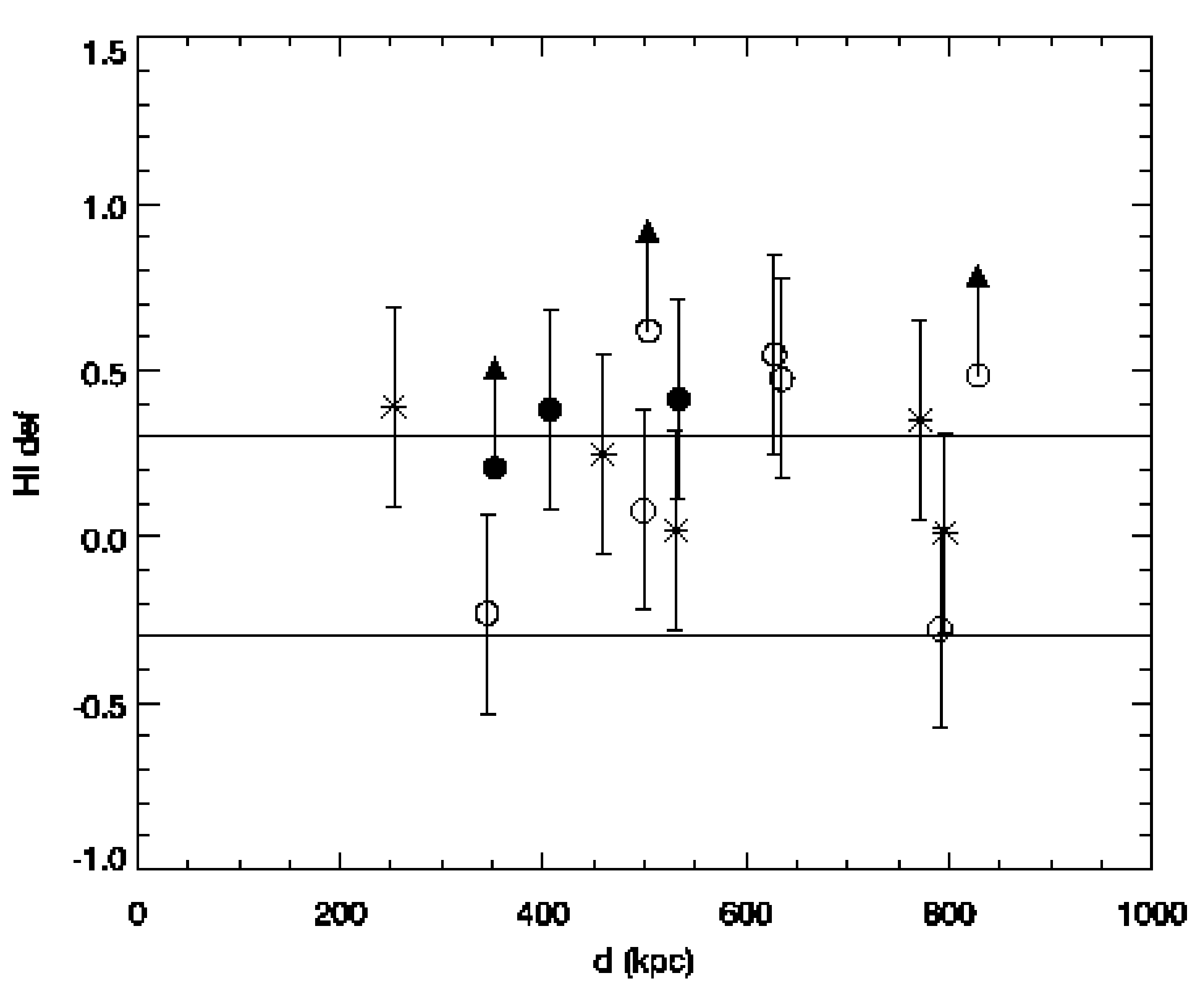}
\caption{The \defhi\ of selected state A, B and C galaxies as a function of distance from the NW-ICM
core.  The following symbols are used to indicate the state (discussed in section \ref{statesnew}) of each spiral: A--(asterisk), B--(open circle) and C--(filled circle).} Galaxies with only upper error bars (arrowheads) are AGES non-detections so the value is the lower limit for \defhi. The region between the horizontal lines
approximates the range expected in the field population.
\label{nw_def}
\end{center}
\end{figure}

To test the model predictions we selected the 14 spirals from our sample which were potentially suffering ram pressure stripping solely attributable to the ICM of the NW
sub--cluster. Virtually all of these galaxies have radial velocities relative to the cluster of $\geq$500 \km\ and it seems reasonable
to assume that most, if not all, are suffering ram pressure
stripping of varying strengths given they are at projected radii of $\leq$\,0.8\,Mpc. Figure \ref{nw_def} shows a plot of  \defhi\ against distance for these galaxies and, interestingly,
only a modest increase in \defhi\ is seen when compared with the field.
Furthermore, no correlation is seen with proximity to the NW sub--cluster ICM core. This observational result
suggests that ram pressure stripping associated with the NW sub--cluster ICM core is relatively weak. The presence of undisrurbed \hi\ in \cg073's dwarf companion projected close to the NW-ICM core provides
additional support to this statement.

Although the  limited evidence for ram pressure stripping is broadly consistent with the expected ram pressure at the projected distances to the the NW sub--cluster core (Figure \ref{pram}),   the lack of correlation with radial distance from the core may imply  that galaxies seen in projection  to be located close to the core  are in reality spread over a  considerable distance. In the case of \cg068, a tidal interaction cannot be ruled out as an explanation of the observed \hi\ characteristics.

\begin{figure}
\begin{center}
\includegraphics[ scale=0.7] {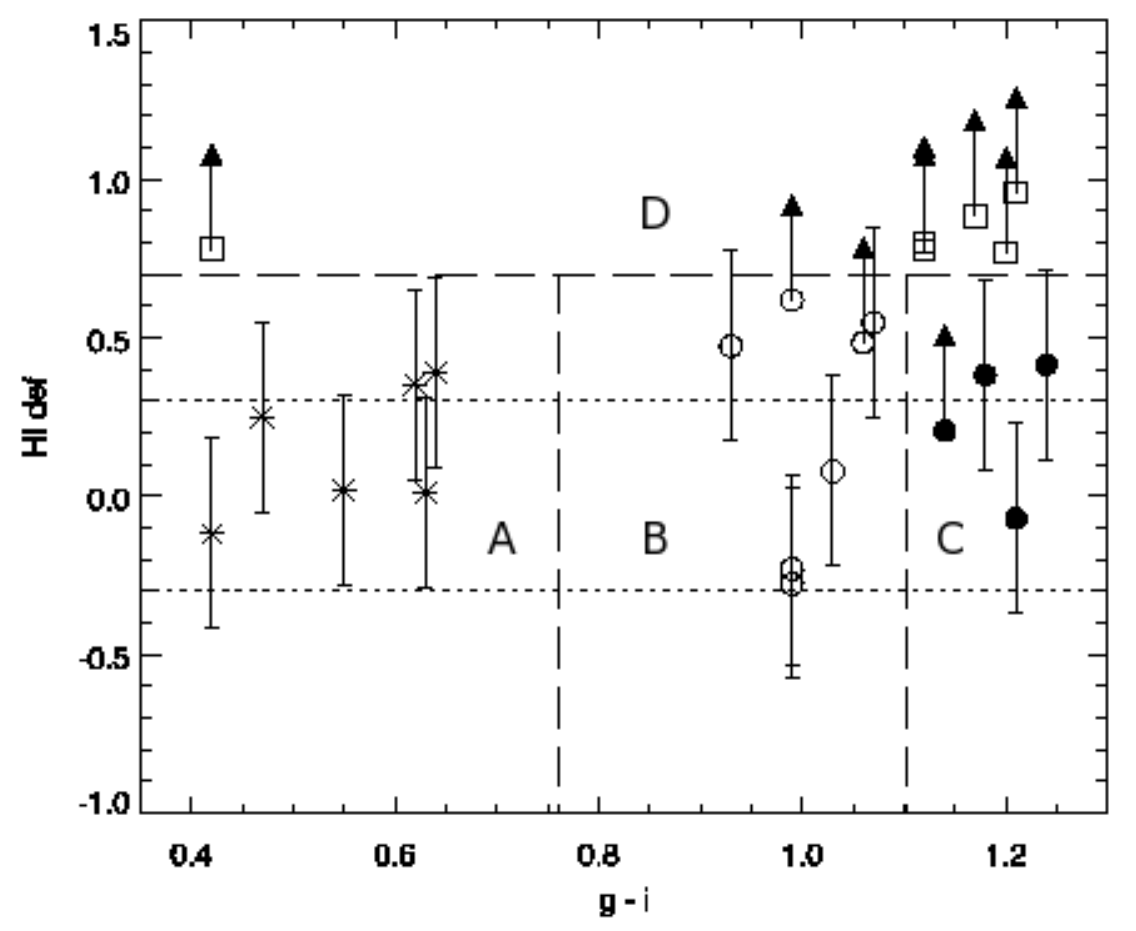}
\caption{The \defhi\ of all the selected late-type galaxies (excluding \cg119 and BIG spirals) plotted against their SDSS \textit{g-i} colour. The long dashed lines mark the boundaries of the parameters defining the four evolutionary states marked A to D. The following symbols are used to indicate the state of each spiral: A--(asterisk), B--(open circle), C--(filled circle) and D--(open square). For AGES non-detections the \defhi\ is a lower limit and is indicated with arrowheads. The region between the dotted horizontal lines approximates the natural range of variation expected in the field population.}
\label{center_def_colour}
\end{center}
\end{figure}

\begin{table*}
\centering
\caption{Evolutionary states of spirals}
\label{ttes}
\begin{tabular}{@{}cllccll@{}}
\hline

 &&&\textbf{Plot}& &\textbf{\hi} &\textbf{\textit{g-i} colour}\\

State &\hi\ &Colour &\textbf{symbol} &number&\textbf{def}& \\
&deficiency &SDSS \textit{g--i}  &  &galaxies&median &median  \\
\hline 

A& $<$0.7&$<$0.76 & $\ast$&6&0.14 &0.59\\ 
B& $<$0.7& 0.76 $<$and $<$1.1 & \hspace{1.25mm}\raisebox{.6ex}{\circle{4.0}}&7&0.47&0.99\\ 
C & $<$0.7& $\geq$1.1 & \textbullet&4&0.30&1.20\\ 
D & $\geq$0.7& any & \framebox(4,4){×}  & 6&0.79&1.15\\ 

\\ \hline
\end{tabular}
\end{table*}
\subsection{Evolutionary states of the spirals}
\label{statesnew}

The colour and  \hi\ deficiency of a spiral are both properties indicative of its evolutionary state. In Figure \ref{center_def_colour} we  distinguish four broad evolutionary states (A--D). We classify the highly \hi\ deficient galaxies (\defhi\ lower limit of $\geq$0.7 ) as state D spirals. The spirals with  moderate or no \defhi\ (i.e., \defhi\ $\leq$0.7) are classified into three states (A, B and C) based on their SDSS \textit{g-i} colour. C state spirals are  red as defined in section \ref{hiages}  (i.e., $g-i$ $\geq$ 1.1). The remaining blue spirals with  moderate or no \defhi\ were classified as state A if their SDSS \textit{g-i} colour was $<$0.76 (the colour of the bluest type of late--type field spirals from \cite{durba08}, their Table 3) and state B for those  with 0.76 $\leq$\textit{g-i} $<$1.1. Table \ref{ttes} summarises the parameters used to determine the evolutionary state of each spiral and the symbols used for them throughout the paper. The evolutionary state determined for each spiral is shown for convenience in the final column of Table  \ref{sdss}. \cg125 and \cg114 have been excluded from the analysis because their evolution is likely to have been dominated by membership of the BIG compact group rather than the cluster.

While the choice of boundaries of our states could be debated, it is clear that the cluster contains spirals at significantly different stages of evolution, with the spirals in each state appearing to have reached a broadly similar evolutionary stage, probably as a result of sharing or passing through a common environment.

The discussion below is focused on the more evolved State C and D spirals where the impacts of the cluster environment are more pronounced. In contrast spirals with  less evolved states A and B  are probably at earlier stages of their interactions and consequently show milder impacts,  although  a key question remains to be addressed, i.e. the mechanism(s) triggering the starbusts (state A).

The \hi\ deficiencies of AGES detected state C spirals are modest compared to lower limits for the state D spirals, suggesting that they have not passed through the cluster's highest density ICM. The state C spirals include 2 of the 3 potentially interacting pairs, where the companion is projected
within 2 arcmin (\aprox 50 kpc), has a velocity within 500 \km\  and both galaxies are $\geq$ 0.5 arcmin (\aprox 12 kpc) in diameter (Table \ref{sdss}). The perturbation parameters, $p_{gg}$, for the state C pairs are \aprox\ 8.7 (\cg102), and \aprox\ 1.8 (\cg111) significantly above the value where the ISM is expected to be impacted. The interaction time scales for the pairs is of the order of 10$^8$ years, in agreement with the value from Boselli $\&$ Gavazzi 2006a. \nocite{bosel06a} It is important to note that   \hi\ emission from close pairs could potentially be confused within Arecibo's \aprox3.5 arcmin beam underestimating the  \defhi\ (AGES) for each member of the pair. But in both cases the spiral's companion is an early type making it unlikely the companion's \hi\ makes a significant contribution.
 Moreover three state C spirals (\cg102, \cg111 and \cg121) have optically disturbed morphologies in SDSS \textit{i}--band images consistent with gravitational interactions. Signs of optical disturbance are also observed in three adjacent  state A spirals, \cg092-1, \cg138 and [BO85]074 showing that tidal interactions, if important, could be affecting spirals in more than one evolutionary state. The high rate of disturbance in the optical morphology in this region suggests the spirals are subject to strong interactions possibly related to their location at the intersection of the NW sub--cluster and the filament oriented toward the Coma
cluster.

The state D spirals all have a red colour, except \cg093 which is exceptionally blue ($g$--$i = 0.42$) and has a disturbed stellar disk. The observed red colours and high \defhi\ are both expected results of a spiral making a transit through a high density ICM core. An example of this is found in Virgo, i.e. the galaxy NGC4568 (Boselli $\&$ Gavazzi 2006a). \nocite{bosel06a} The calculations for the NW sub--cluster (section \ref{rpsnw}) suggests that a core transit would be required to cause the \hi\ deficiencies observed in the state D spirals under the ram pressure stripping scenario, although the ICM core transited may have been the SE subcluster. The tendency for the state D spirals further from the cluster centre to be redder and have higher velocities may be indicating that these spirals are a  population of \textit{backsplash} galaxies \citep{elling04},  which were ram pressure stripped as they passed through the SE sub--cluster ICM core with a significant velocity component directed to the west.\\

\section{concluding remarks}
\label{conclusion}

We have presented medium--resolution VLA data for a field located to the NW of A\,1367's centre, and some higher--resolution, but less sensitive, VLA data for a field containing BIG. By using a catalogue of the brightest late-type
members of the cluster,  combining AGES and SDSS data, we have identified four evolutionary states with shared characteristics in terms of their combination of \hi\ content and colour.

The overall  picture from the HI deficiencies as a function of radial distance and calculations of ram pressure for the NW cluster  (section \ref{rpsnw}) are both consistent with moderate levels of ram pressure stripping. However the presence of spirals with a range of evolutionary states in projected and velocity space, particularly the  spirals with more advanced evolutionary states (state C and D spirals), suggests an evolutionary process or processes operating on relatively local scales.

Consideration of the membership of the evolutionary states and individual spirals presents mixed, and some times contradictory, indicators of varying reliability for the dominant ISM removal process(es). The lack of correlation between \defhi\ and proximity to the NW sub--cluster core for the spirals likely to be affected by it, the presence of starbusts, the presence of galaxies with \hi\ offset directions inconsistent with ram pressure,  together with the elevated number of morphologically disturbed red and moderately \hi\ deficient  spirals (state C) possibly linked to interactions during infall, all suggest tidal effects. On the other hand the presence of the highly \hi\ deficient (state D) spirals and the possible case of gas falling back into the galaxy's potential following an ISM core transit (\cg072) support the ram pressure stripping model.

While the analysis above hints that different ISM removal mechanisms may be active in this cluster further observations are required to discriminate between the hydrodynamic and tidal mechanisms. If anything, this paper shows that the picture in Abell 1367 is somewhat confusing, probably the result of it being less evolved than Coma and Virgo.

\section*{Acknowledgments}

We thank the referee for comments to an earlier version of the manuscript which have led to a much improved presentation.

HBA thanks CONACyT for grant No.\ 50794, the Royal Astronomical Society and the Academia Mexicana de Ciencias for financial support to carry out this work. CAC acknowledges the University of Guanjuato (DINPO) for grant 0131/07. MJH thanks the Royal Society for a Research Fellowship.

This research has made use of the NASA/IPAC Extragalactic Database (NED) which is operated by the Jet Propulsion Laboratory, California Institute of Technology, under contract with the National Aeronautics and Space Administration.

This research has made use of the Sloan Digital Sky Survey (SDSS). Funding for the SDSS and SDSS-II has been provided by the Alfred P. Sloan Foundation, the Participating Institutions, the National Science Foundation, the U.S. Department of Energy, the National Aeronautics and Space Administration, the Japanese Monbukagakusho, the Max Planck Society, and the Higher Education Funding Council for England. The SDSS Web Site is http://www.sdss.org/.

This work is partly based on observations obtained with {\it
XMM-Newton}, an ESA science mission with instruments and contributions
directly funded by ESA Member States and the USA (NASA).

\bibliographystyle{mn2e}
\bibliography{cluster}

\newpage
\begin{appendices}
	\begin{center}
		\textbf{\appendixpage}
	\end{center}
\appendix
\section{\textbf{BIG}}

The Blue Infalling Group is an interesting case of a compact group that has been suggested to be infalling toward the core of the SE sub--cluster with a relative velocity
of \aprox 2000 \km\ \citep{cort06}. As well as the two principal galaxies
\cg114 and \cg125, BIG contains an extraordinarily high concentration of
dwarf galaxies with high star formation rates and extragalactic \hii\ regions
\citep[Gavazzi \al 2003a]{sakai02}.\nocite{gava03a} Deep \halpha\ imaging has revealed extensive gas
streamers interpreted as resulting from a strong tidal interaction between
members of the group \citep{cort06}. \hi\ in BIG has previously been imaged
using the Westerbork Synthesis Radio Telescope (WSRT) by \citet{sakai02}.

Our observations have higher spatial resolution that clearly show the offset between the highest density \hi\ and the optical galaxy  (Figure \ref{big125}). Weak \hi\ emission was detected by us at the position of a knot 15 arcsec (6 kpc)
east of knot K2a, with peak \hi\ emission at 8190 \km\ or $ \sim $70 \km\ greater
than K2a's optical velocity (Figure \ref{big125}). However, we did not
pick up the large continuous region of low density \hi\ extending west of
\cg125 and running from below K2a north to DW2 seen with the WSRT
\citep{sakai02}, which is probably due to our lower sensitivity. We were also
unable to confirm in our robust 0 cube the \hi\ to the west of the optical position
of the irregular blue galaxy \cg114, observed with the WSRT, although weak
emission was found after applying natural weighting. The weak detection of
\cg114 implies a large \hi\ deficiency which is confirmed by the Arecibo
determined \hi\ mass of 3 x 10$^8$ \msolar\ \citep{cort06}.

Regarding the relative location of BIG with respect to the clus-
ter centre, this is still matter of debate (Sakai et al. 2002; Gavazzi 
et al. 2003a; Cortese et al. 2006) \nocite{sakai02,gava03a,cort06}.
The calculation discussed in section 5.2 indicates that, if
the group is within $\sim$ 0.5 Mpc from the cluster centre, we should expect
evidence of strong ram pressure stripping.
This, however, would seem to be contradicted
by the presence of substantial amounts of extended H I in the velocity range from 8000 to 8500 \km ,   suggesting
either that BIG lies in the background of A1367 or it has just
started its infall into the cluster center.

\newpage
\section{\textbf{ABELL 1367[GP82]1227}}

\begin{figure}
\begin{center}
\includegraphics[scale=0.45] {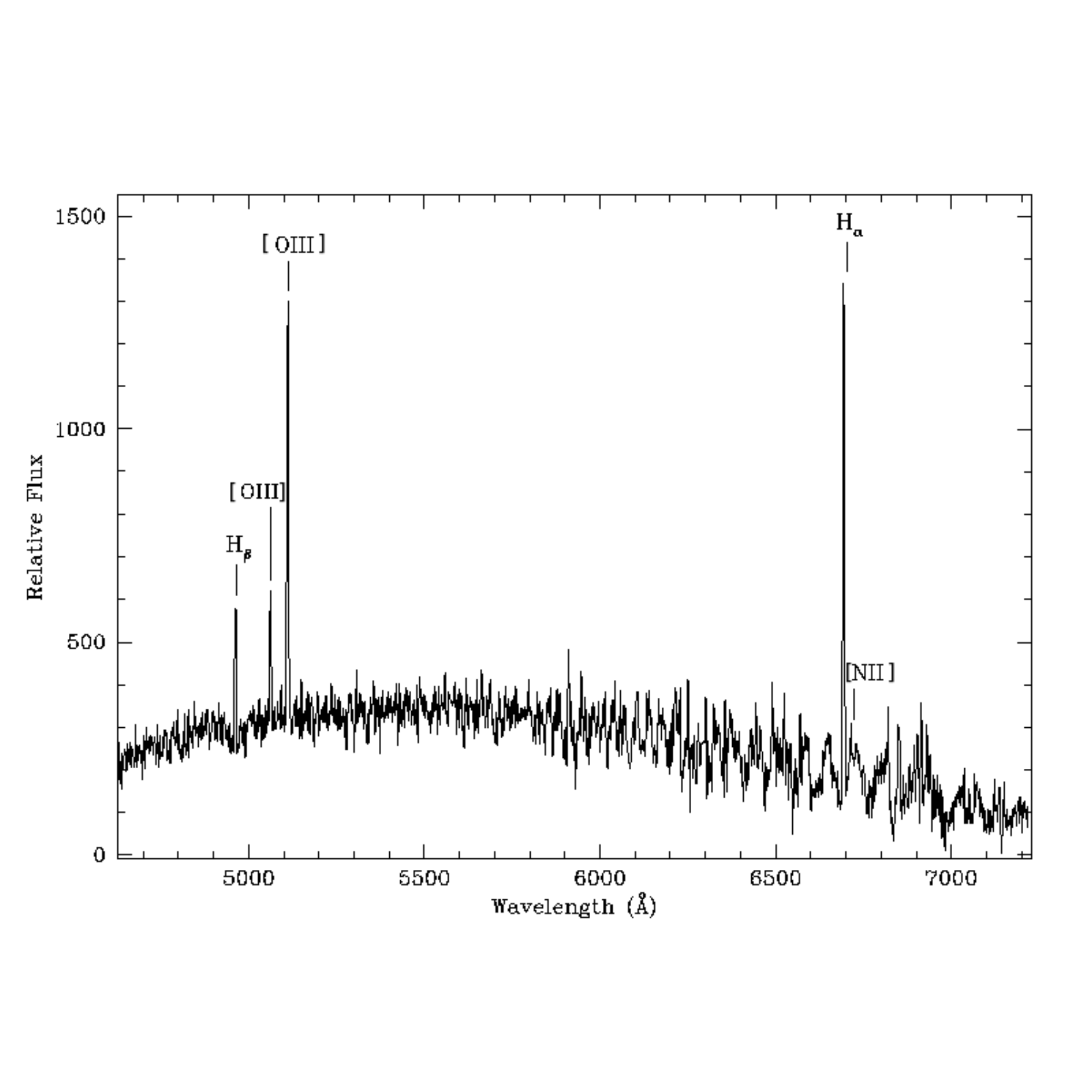}
\caption{{\bf A\,1367\,[GP82]1227} emission line spectrum: SPM
emission line spectrum showing $H_{\beta}$, [OIII] (4958.9), [OIII]
(5006.8) [NII](${\lambda}$ 6584) and $H_{\alpha}$. }
\label{1227_r}
\end{center}
\end{figure}

We obtained a low resolution (4.8 \AA) spectrum of A\,1367\,[GP82]1227,
covering the wavelength range 4630-7225 \AA, in May 2007 (Figure
\ref{1227_r}). The observation was carried out with the 2.12--m telescope at
the San Pedro M\'artir Observatory (SPM) in Mexico, using a Boller \& Chivens
spectrograph, a 600 l/mm grating and a 2K Thomson CCD. This configuration
gave a dispersion of 1.3 \AA\ per pixel. The slit was placed in an East-West
direction passing through the central region of the galaxy. Four 20 minute
exposures were taken and the spectra were averaged after cosmic
ray cleaning, extraction and wavelength calibration. A wavelength solution
using about 30 He+Ar+Ne lamp lines, gave an uncertainly of \aprox 0.18 \AA\
rms. Sky transparency was good, with the seeing slightly above 1 arcsec. Due
to the faintness of the galaxy, the signal to noise obtained was only about 5
in the range 5500 - 6500 \AA.

The spectrum of A\,1367\,[GP82]1227 shows the emission lines of H$_{\beta}$ (${\lambda}_\mathrm{rest}$ = 4861.3), [OIII]
(4958.9), [OIII] (5006.8) and H$_{\alpha}$ (6562.8) at 4963.8,
5063.2, 5112.3 and 6700.4 \AA, respectively. This gives an emission--line
radial velocity of 6307 \km, which is in good agreement with our \hi\
velocity of 6239 \km\ (see Table \ref{table2}). These emission lines indicate active star
formation in A\,1367\,[GP82]1227. The [OIII](${\lambda}$ 5007)/H$_{\beta}$
and [NII](${\lambda}$ 6584)/H$_{\alpha}$ ratios are 2.64 and 0.03,
respectively (note that [NII] is barely visible in the spectrum) putting this
galaxy completely in the locus of star forming galaxies in a BPT
diagram \citep{btp81}. It was not possible to resolve the \halpha\ regions individually
with the SPM 2.12m.

\newpage
\section{\textbf{CHANNEL MAPS OF THE VLA DETECTIONS}}
The channel maps for all  VLA  detections are available as supplementary online material. Below as an example of the online material are the  channel maps for \cg079.
\clearpage
\newpage

\begin{figure*}
\begin{center}
\includegraphics[scale=0.75] {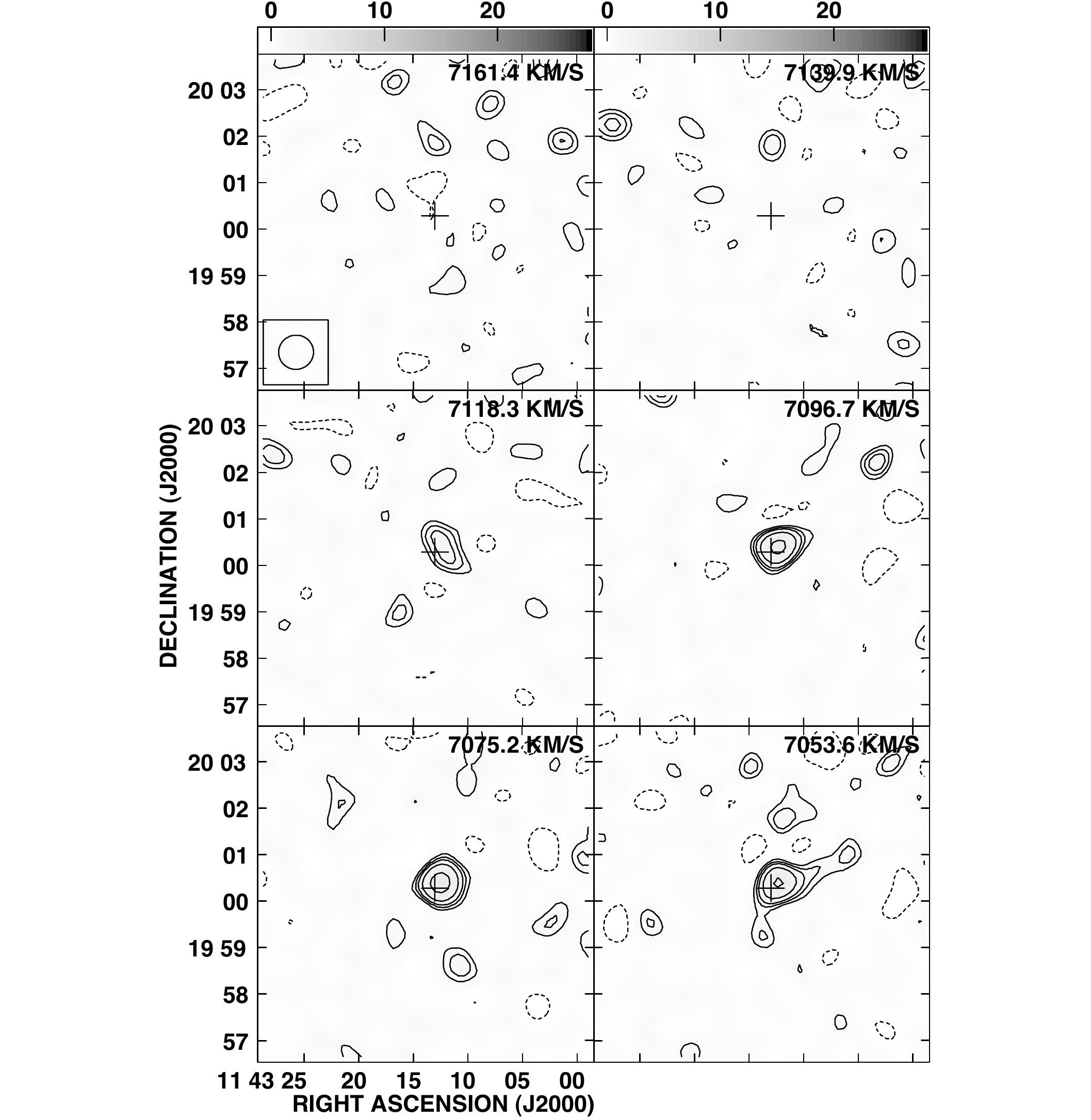}
\caption{ Channel maps for\textbf{ \cg079.} \first 0.36 mJy. \last   }
\label{97079a_cm}
\end{center}
\end{figure*}

\clearpage
\newpage
\addtocounter{figure}{-1}
\begin{figure*}
\begin{center}
\includegraphics[scale=0.75] {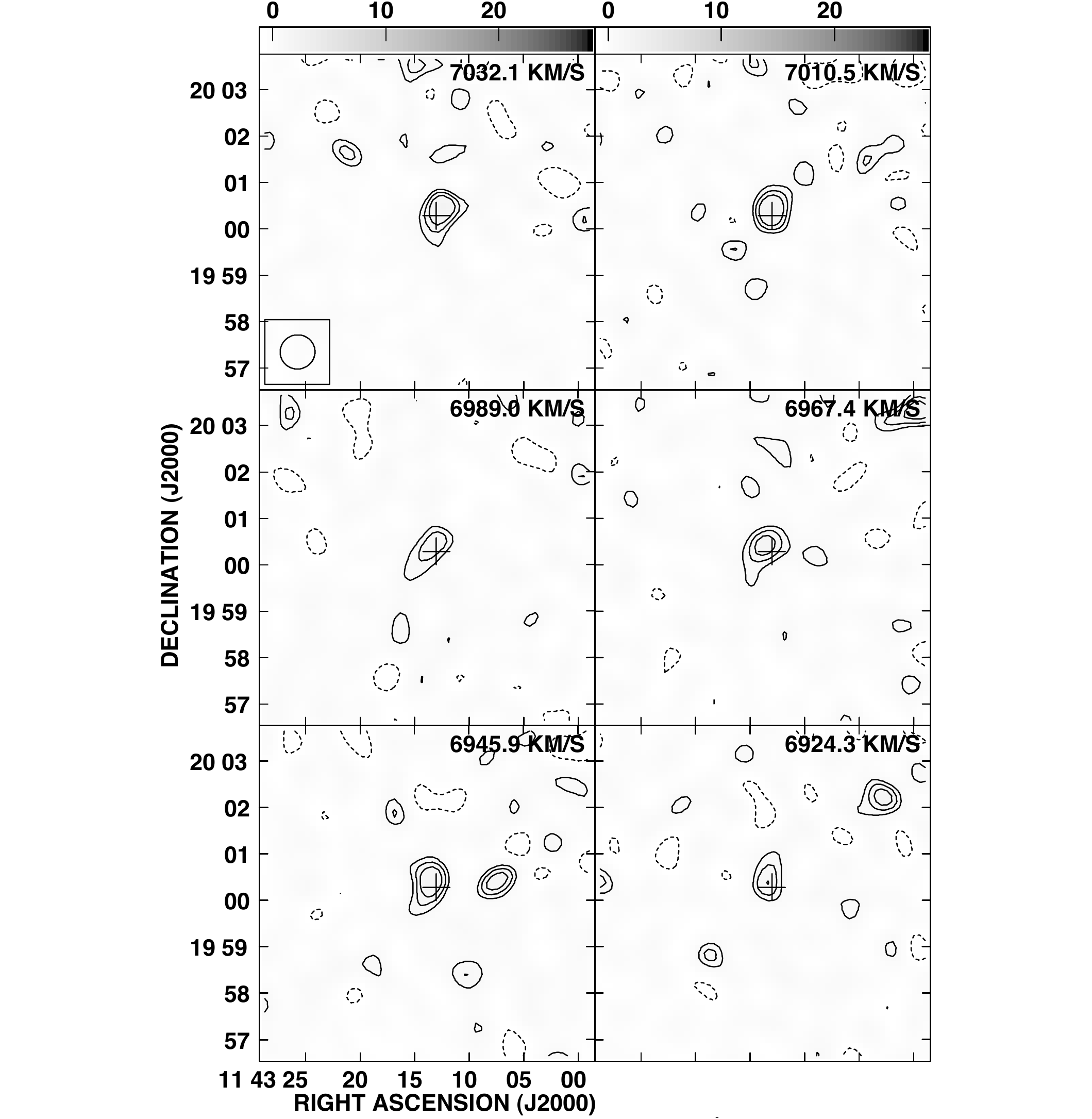}
\caption{-- \textit{continued} }
\label{97079b_cm}
\end{center}
\end{figure*}

\clearpage
\newpage
\addtocounter{figure}{-1}
\begin{figure*}
\begin{center}
\includegraphics[scale=0.75] {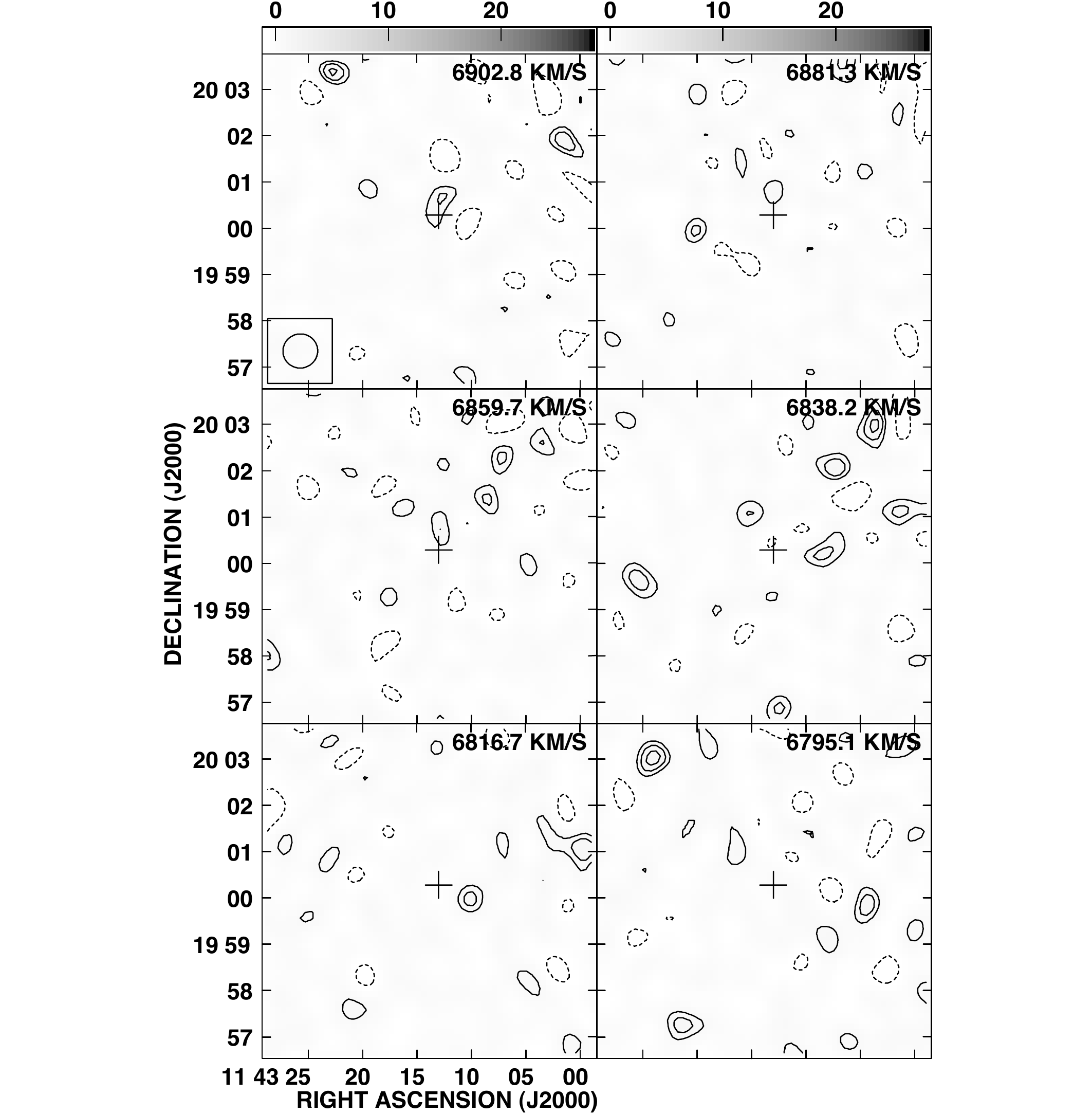}
\caption{-- \textit{continued} }
\label{97079c_cm}
\end{center}
\end{figure*}

\end{appendices}

\end{document}